\documentclass[a4paper, 11pt]{article}
\pdfoutput=1

\usepackage{jcappub}
\usepackage[utf8]{inputenc}
\usepackage{graphicx}
\usepackage{amsmath}
\usepackage{amsfonts}
\usepackage{amssymb}
\usepackage{bbold}
\usepackage{float}

\usepackage{tikz}
\usetikzlibrary{arrows,backgrounds,snakes,patterns}
\usetikzlibrary{shapes,arrows,chains}
\usepackage{verbatim}
\usepackage{booktabs}
\usepackage{pgfplots}
\pgfplotsset{compat=1.10}
\usepgfplotslibrary{fillbetween}
\usepackage[flushleft]{threeparttable}

\usepackage{array}
\usepackage{placeins}

\usepackage{subcaption}

\newcommand{\cG}{{\cal G}}
\newcommand{\cD}{{\cal D}}
\newcommand{\cM}{{\cal M}}
\newcommand{\cR}{{\cal R}}
\newcommand{\df}{\dot{\phi}}
\newcommand{\ds}{\dot{\sigma}}
\newcommand{\Mpl}{\ensuremath M_{\textrm{Pl}{}}}

\numberwithin{equation}{section}
\numberwithin{figure}{section}

\dedicated{UTTG-05-19}
\title{Multi-field Inflation in High-Slope Potentials}

\author[a]{Vikas Aragam}
\author[a]{Sonia Paban}
\author[a]{Robert Rosati}

\affiliation[a]{
	Theory Group, Department of Physics, University of Texas at Austin, Austin, TX 78712, USA
}

\emailAdd{aragam@utexas.edu}
\emailAdd{paban@physics.utexas.edu}
\emailAdd{rjrosati@utexas.edu}

\abstract{
We present two families of multi-field potentials that support inflation while satisfying the refined de Sitter and the distance swampland conjectures. Both families feature Planck-compatible phenomenology. The first is a helix-type potential, in a flat field-space metric, that satisfies the conjectures via a high turning rate. This model has a tensor-to-scalar ratio close to, but below, the current experimental limits and small non-gaussianities. The second family, an example of orbital inflation, utilizes a negatively curved field metric to achieve prolonged inflation with nontrivial turning in the presence of a tachyonic direction. Although perturbations in this model undergo an exponential growth before horizon exit, it is always possible to match the measured amplitude of the power spectrum by lowering the scale of inflation if the turning rate is low enough. We identify a Planck-compatible region of parameter space in which the scale of inflation is above that of nucleosynthesis. Due to the rapid growth, this model predicts an exponentially suppressed value for the tensor-to-scalar ratio.
}
\arxivnumber{1905.07495}
\begin{document}
	
\maketitle
\thispagestyle{empty}
\newpage

\setcounter{page}{1}

\section{Introduction}

In multi-field models it is possible to have both a period of inflation and a steep potential \cite{Yang:2012bs, Achucarro:2018vey, Brown:2017osf} \footnote{See \cite{Copeland:2000hn, Sami:2003my, Anber:2009ua, Adshead:2012kp, Anber:2012du, Rezazadeh:2015dia,Adshead:2016iix,Dimopoulos:2018upl} for exploring alternative ways of making inflation compatible with steep potentials. }.
In these models the ratio of the gradient of the potential to the potential is still proportional to the rate of time evolution of the Hubble parameter, but the proportionality coefficient can be much larger than one: 

\begin{equation}
\epsilon_V \simeq \epsilon_H \left(   1 + \frac{\omega^2}{9 H^2}     \right),  \hspace{3ex} \epsilon_V= \frac{M_{Pl}^2}{2} \left( \frac{|\nabla V| }{V}\right)^2, \hspace{3ex}  \epsilon_H= - \frac{\dot{H}}{H^2}.
\label{releps}
\end{equation}
Here $\omega$  (called $\Omega$ in some references) measures the turning rate of  the trajectory, and it has been assumed that $\epsilon_H \, \rm{and} \, |\eta_H| \ll 1$.  The relation (\ref{releps}) is a statement about the background field motion. It is the first step toward a viable inflation model, but  fitting the experimental data will impose further constraints both on the classical trajectory (number of e-folds) and on the quantum fluctuations: $n_s$, $r$, isocurvature power and non-gaussianities. It is known that models of  multi-field inflation  can produce sizable isocurvature power  whenever $\omega \neq 0$ for some field masses  \cite{Wands:2007bd, Cremonini:2010sv, Cremonini:2010ua, Achucarro:2010da}. It is also known that they can  be unstable \cite{Renaux-Petel:2015mga}. For these reasons it is important to build and analyze actual models. 

In this paper we  construct potentials with large values of $\epsilon_V  \gtrsim 1 $, both in  flat and negatively curved field-space geometries, and probe their compatibility with the experimental limits. More precisely, the examples below satisfy the constraint \cite{Agrawal:2018own} 
\begin{equation} \label{eq:revswamp}
\begin{aligned}
&\hspace{9ex}\epsilon_V \gtrsim 1 \hspace{3ex} \text{or} \hspace{3ex} \eta_V \lesssim -1 \\ \\
&\eta_V \equiv \Mpl^2 \times \text{minimum eigenvalue}( V_{;IJ}) / V
\end{aligned}
\end{equation}
in a region of field space of magnitude $ \Delta \phi \gtrsim \Mpl$, where $\Delta \phi$ is the geodesic distance in field-space. The two examples we provide will illustrate that imposing compatibility with CMB data significantly reduces the parameter space that satisfies (\ref{releps}).   The possibility of generating inflation in  potentials with these properties is intriguing in view of the swampland conjectures \cite{Obied:2018sgi,Ooguri:2018wrx, Garg:2018reu, Garg:2018zdg,Agrawal:2018own,Andriot:2018mav}\footnote{For a follow-up on the cosmological consequences of the conjectures, see \cite{
Rudelius:2019cfh,
Das:2019vnx,
Benetti:2019smr,
Colgain:2019joh,
Sabir:2019bsh,
Rajvanshi:2019wmw,
CaboBizet:2019sku,
Micu:2019fju,
Ijjas:2019pyf,
vandeBruck:2019vzd,
Brahma:2019mdd,
Mukhopadhyay:2019cai,
Colgain:2019pck,
Slosar:2019flp,
Kaloper:2019lpl,
Gonzalo:2019gjp,
Sabir:2019wel,
Farakos:2019ajx,
Christodoulidis:2019jsx,
Brahma:2019kch,
Bjorkmo:2019fls,
Berglund:2019pxr,
Almeida:2019iqp,
Lynker:2019joa,
Kadir:2010dh,
Heisenberg:2019qxz,
Fumagalli:2019noh,
Artymowski:2019vfy,
Carta:2019rhx,
Kamali:2019hgv,
Nan:2019gjw,
Heckman:2019dsj,
Bramberger:2019zez,
Chway:2019prm,
Bjorkmo:2019aev,
Kobakhidze:2019ppv,
Achucarro:2019pux,
Kamali:2019ppi,
Draper:2019zbb,
Arciniega:2018tnn,
Belgacem:2018wtb,
Cai:2018ebs,
Raveri:2018ddi,
Coudarchet:2018ezq,
Abel:2018zyt,
Brax:2018hyw,
Seo:2018abc,
Scalisi:2018eaz,
Bastero-Gil:2018yen,
Bhardwaj:2018omt,
Gonzalo:2018guu,
Lin:2018edm,
Montero:2018fns,
Heisenberg:2018erb,
Hertzberg:2018suv,
Kinney:2018kew,
Bonnefoy:2018mqb,
Herdeiro:2018hfp,
Acharya:2018deu,
Banlaki:2018ayh,
Emelin:2018igk,
Holman:2018inr,
Tosone:2018qei,
Elizalde:2018dvw,
Cheong:2018udx,
Thompson:2018ifr,
Chiang:2018lqx,
Heckman:2018mxl,
Yi:2018dhl,
Agrawal:2018rcg,
Kim:2018mfv,
Lin:2018rnx,
Schimmrigk:2018gch,
Dvali:2018jhn,
Garg:2018zdg,
Hebecker:2018vxz,
Fukuda:2018haz,
Wang:2018kly,
Ooguri:2018wrx,
Das:2018rpg,
Antoniadis:2018ngr,
Ashoorioon:2018sqb,
Odintsov:2018zai,
Motaharfar:2018zyb,
Kawasaki:2018daf,
Hamaguchi:2018vtv,
Lin:2018kjm,
Draper:2018lyw,
Benisty:2018fja,
Dimopoulos:2018upl,
Matsui:2018xwa,
Visinelli:2018utg,
Han:2018yrk,
DAmico:2018mnx,
Brandenberger:2018wbg,
Wang:2018duq,
Danielsson:2018qpa,
Das:2018hqy,
Quintin:2018loc,
VanDenHoogen:2018anx,
Choi:2018rze,
Brahma:2018hrd,
Marsh:2018kub,
Murayama:2018lie,
Heisenberg:2018rdu,
Akrami:2018ylq,
Cicoli:2018kdo,
delRio:2018vrj,
Dasgupta:2018rtp,
Dutta:2018vmq,
Kinney:2018nny,
Gu:2018akj,
Damian:2018tlf,
Heisenberg:2018yae,
Chiang:2018jdg,
Ben-Dayan:2018mhe,
Matsui:2018bsy,
Andriot:2018ept,
Roupec:2018mbn,
Ghalee:2018qeo,
Paban:2018ole,
Colgain:2018wgk,
Denef:2018etk,
Dias:2018ngv,
Rasouli:2018kvy,
Kehagias:2018uem,
Lehners:2018vgi,
Garg:2018reu,
Dimopoulos:2018eam,
Achucarro:2018vey,
Aalsma:2018pll,
Heisenberg:2018vsk,
Banerjee:2018qey,
Dvali:2018fqu,
Andriot:2018wzk,
Agrawal:2018mkd,
Obied:2018sgi,
Vagnozzi:2018jhn,
Visinelli:2017imh,
Brahma:2019iyy}. An interesting possibility is that the conjectures are a consequence of forbidding eternal inflation \cite{Rudelius:2019cfh}.}, though the approach followed in this paper is bottom-up. In fact, we lack a compelling reason to expect a top-down approach to generate the combination of metric and potential of the examples that we analyze.

 Hyperinflation \cite{Brown:2017osf}  is an interesting idea that balances large potential gradients against the (negative) curvature of field space to generate a period of inflation. Negative curvature field-space metrics appear frequently in string theory compactifications. The second family of solutions we present in this work uses the same field-space metric as Hyperinflation but a different potential.  The follow-up work to the initial Hyperinflation proposal has focused on its quantum fluctuations  \cite{Mizuno:2017idt}.  In particular, \cite{Renaux-Petel:2015mga, Renaux-Petel:2017dia, Cicoli:2018ccr, Grocholski:2019mot} have pointed out that the perturbations experience an exponential growth before horizon exit due to the rapid turning trajectory and the negatively curved field space. This is not fatal as long as the turning rate is low enough \cite{Fumagalli:2019noh, Bjorkmo:2019qno}.  Further work by Bjorkmo and Marsh \cite{Bjorkmo:2019aev, Bjorkmo:2019fls} has generalized the idea  of hyperinflation to models with more than two fields and a broader class of potentials.
 
The two families of potentials analyzed in this paper are distinct from hyperinflation, but  create similarly strong non-geodesic inflationary trajectories. The helix-type potential  assumes a flat field space metric and is hence clearly different from hyperinflation.  The second family is a particular example of orbital inflation \cite{Achucarro:2019pux, Christodoulidis:2019mkj} which assumes the same negatively curved field space as hyperinflation but has a different potential. Its perturbations face similar challenges to hyperinflation.  As we will explain in  section 4, this model has a region of parameter space for which the refined swampland constraints are compatible with current experimental data. As in hyperinflation, these solutions lead only to experimentally compatible results when inflation happens at very low scales.  A measurement of the metric perturbations  in the near future would rule out this family of models as realizations of inflation that are compatible with the refined swampland conjectures.

\section{Notation}
Before describing our models, a quick note on notation.
We consider models with $N_f$ scalar fields in $(3+1)$  spacetime dimensions and Friedman-Lema\^{i}tre-Robertson-Walker metric, with spacetime metric signature $(-, +,+,+)$. The field-space has a metric $\cG_{IJ} (\phi^I)$. Greek letters label space-time indices, lower-case Latin letters label spatial indices and upper-case Latin letters label field-space indices, $I,J = 1,2,\ldots,N_f$. We work in units where the reduced Planck mass is set to one, but occasionally insert it in expressions to make dimensions apparent. 

With these assumptions the equations of motion for the background fields are:

\begin{equation}\label{eq:multiFieldEoM}
\cD_t \dot{\phi^I} + 3 H \dot{\phi^I} + \cG^{IJ} V_{,J}=0 \\
\end{equation} where $V_{,I} \equiv \cD_I V$. The covariant derivative with respect to cosmic time is defined as:
\begin{equation}
\cD_t A^I \equiv \dot{\phi}^J \cD_J A^I = \dot{A}^I + \Gamma^I_{JK} \, A^J \df^K .
\end{equation}
As these two equations show, it is possible to offset large gradients in the potential against curvature to have slow-roll inflation.

\subsection{Perturbations}
This section largely follows the notation of \cite{Kaiser:2012ak}.  The evolution of the perturbations is given by:

\begin{equation}
\cD_t^2 Q^I + 3 H \cD_t Q^I + \left[   \frac{k^2}{a^2} \delta^I_J + \cM^I_J - \frac{1}{a^3} \cD_t \left(\frac{a^3}{H} \dot{\phi}^I \dot{\phi}_J \right)          \right] Q^J =0
\label{eq:perteom}
\end{equation}
where $Q^I$ are the Mukhanov-Sasaki variables. They are gauge invariant with respect to space-time gauge transformations to first order in the perturbations.  The mass-squared matrix appearing in the equation of motion for the perturbations is

\begin{equation}\label{eq:massMatrix}
{\cM^I}_J \equiv \cG^{IK} ( \cD_J \cD_K V) - \cR^I_{LMJ} \df^L \df^M 
\end{equation} where $\cR^I_{LMJ}$ is the Riemann tensor for the field-space manifold. We may decompose the perturbations along directions tangent (adiabatic: $Q_\sigma$) and perpendicular (entropic: $\delta s^I$) to the classical trajectory:

 \begin{eqnarray*}
 Q_{\sigma} &\equiv & \hat{\sigma}_I Q_I = \frac{ \dot{\sigma}}{H} \cR_c, \hspace{3ex} \mathrm{where} \hspace{3ex}  \hat{\sigma}^I \equiv \frac{\df^I}{\ds}, \hspace{3ex} \dot{\sigma}^2 \equiv \cG_{IJ} \dot{\phi}^I \dot{\phi}^J\\
 \delta s^I & \equiv& {\hat{s}^I}_J Q^J, \hspace{3ex} \hat{s}^{I J} \equiv \cG^{IJ} -\hat{\sigma}^I \hat{\sigma}^J \\
  \omega^I & \equiv & \cD_t \hat{\sigma}^I=- \frac{1}{\dot{\sigma}} V_{,K}   \hat{s}^{I K}\hspace{3ex} \omega=|\omega^I|.
\end{eqnarray*}  Here, $\cR_c$ is the gauge invariant curvature perturbation. The equation for the adiabatic  mode is:

\begin{equation}\label{eq:adiabaticEOM}
\ddot{Q}_{\sigma} + 3 H \dot{ Q_{\sigma}} + \left[       \frac{k^2}{a^2} + \cM_{\sigma \sigma} - \omega^2 -\frac{1}{ a^3 } \frac{d}{dt} \left( \frac{a^3 \dot{\sigma}^2}{H} \right)         \right] Q_{\sigma}= 2 \frac{d}{dt} (\omega_J \delta s^J) - 2 \left( \frac{V_{, \sigma} }{\dot{\sigma}}+\frac{\dot{H}}{H}\right)  (\omega_J \delta s^J).
\end{equation} This indicates that there is a particular combination of entropic modes with special physical significance $(\omega_J \delta s^J)$. To separate this combination from the rest, one introduces a unit vector that points in the direction of the turning rate:

\begin{eqnarray}
\hat{s}^I &\equiv& \frac{\omega^I}{\omega},   \hspace{4ex} \gamma^{IJ} = G^{IJ} -\hat{\sigma}^I \hat{\sigma}^J - \hat{s}^I \hat{s}^J \label{eq:defsg} \\
\delta s^I &=& \hat{s}^I Q_S + B^I  \hspace{3ex} {\rm{where}} \hspace{3ex} Q_s \equiv \hat{s}_J Q^J,  \hspace{3ex} B^I \equiv {\gamma^I}_J Q^J.
\end{eqnarray}
The evolution of $Q_s$ is given by the equation:

\begin{eqnarray}
\ddot{ Q}_{s} &+& 3 H \dot{ Q}_{s} + \left[       \frac{k^2}{a^2} + \cM_{ss} + 3 \omega^2 -\Pi^2         \right] Q_s \label{eq:entropicEOM} \\ 
&= &4 \frac{k^2}{a^2} \frac{\omega}{\dot{\sigma}} \Psi  -\cD_t ( \Pi_J B^J) - \Pi_J \cD_t B^J -\cM_{IJ} \hat{s}^I B^J - 3H ( \Pi_J B^J) \label{eq:entroevol} 
\end{eqnarray}
where $\Pi^I =\frac{1}{\omega} \cM_{KJ} \hat{\sigma}^K  \gamma^{IJ}$, $\cM_{ss} \equiv \cM_{IJ}  \hat{s}^I \hat{s}^J$, and
\begin{align}
	\frac{k^2}{a^2}\Psi = \frac{\dot{H}}{H}\left[\frac{d}{dt}\left(\frac{H}{\dot{\sigma}}Q_\sigma \right) - \frac{2H \omega}{\dot{\sigma}}Q_s \right].
\end{align}
Taking the sub- and super-horizon limits respectively, the evolution of $Q_s$ becomes:
\begin{align}\label{eq:entropicEomSubSuper}
0 = \ddot{Q_s} + 3H\dot{Q_s} + 
\begin{cases}
\left(\frac{k^2}{a^2} + \mathcal{M}_{ss} - \omega^2 \right)Q_s - 4\frac{\omega}{\dot{\sigma}}\frac{\dot{H}}{H}\frac{d}{dt}\left(\frac{H}{\dot{\sigma}} Q_\sigma \right) & \text{if } k^2 \gg (aH)^2 \\
\left(\mathcal{M}_{ss} + 3\omega^2 \right)Q_s & \text{if } k^2 \ll (aH)^2
\end{cases}
\end{align}
We denote the entropic mode's effective super-horizon mass in two-field inflation as $\mu_s^2 \equiv \cM_{ss} + 3 \omega^2$ and the effective sub-horizon mass as $\mu_{s,sub}^2 \equiv \mathcal{M}_{ss} - \omega^2$.

\section{Helix-trajectory potentials}
\label{sec:helix}
\par 

In this section we present a class of three-field helix-like potentials in flat field-space with a high turning rate, a large $\epsilon_V$, and observationally consistent phenomenology.

Different potentials with helix-type behavior have been studied before in \cite{Berg:2009tg,Barenboim:2014vea,Achucarro:2019pux}.  However, all of them are two-field models and none of them support high-slope inflation with $\epsilon_V \gtrsim \mathcal{O}(1)$. 
Other differences include: trajectories following the minima of Dante's Inferno \cite{Berg:2009tg} produce no turning. Spiral Inflation produces turning \cite{Barenboim:2014vea} but has a tachyonic mode and a growing radius. The potential presented below is single-valued while the example of Shift-symmetric Orbital Inflation in \cite{Achucarro:2019pux} is multi-valued and has an additional shift symmetry which guarantees a flat direction in the potential. We later analyze an instance of orbital inflation in section \ref{sec:super}.

\begin{figure}
\centering
\includegraphics[width=0.6\textwidth]{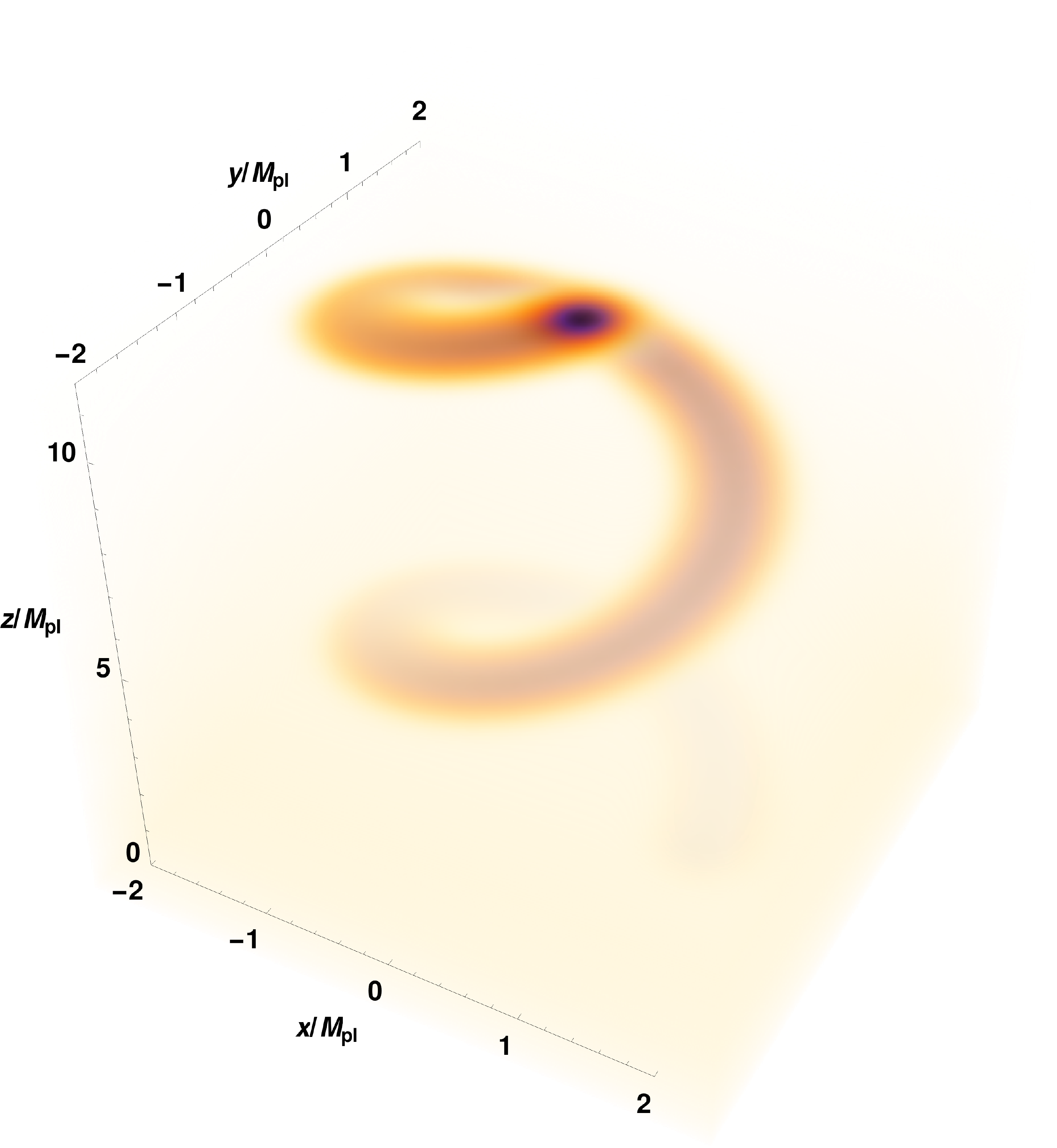}
\caption{The helix-path potential (\ref{eq:helixpotential}). The potential is plotted so that cloud density increases as $V$ decreases. Note that the center of the helical track is always a constant $\Delta$ lower than its surroundings, while the decrease in the $e^{z/R}$ term appears as an increasing cloudiness in the volume around the helix towards the bottom of the plot, partially hiding it from view. We encourage readers to view this plot in color. Parameters chosen for this plot were illustrative and not in the regime of inflationary interest.}
\label{fig:helixpicture}
\end{figure}

The helical potential described here is the first flat field space construction to locally satisfy the dS conjecture and produce observationally consistent phenomenology.
This potential forces a helical trajectory in field space. There are three fields, $x,y,z$, with canonical kinetic terms.
\begin{align}
V = \Lambda^4 \left( e^{z/R} + \Delta \left(1 - \exp\left[\frac{-(x - A \cos{z/f})^2 - (y - A \sin{z/f})^2)}{
         2 \sigma^2}\right]\right)\right)
\label{eq:helixpotential}
\end{align}
The potential is exponential in $z/R$, other than a gaussian divot curled into a helix with radius $A$ and period $2\pi/f$. The depth and width of the divot are set by $\Delta$ and $\sigma$ respectively. See figure \ref{fig:helixpicture}.

This potential can fulfill the dS-conjecture in a wide inflationary region.
It will be helpful to define new fields $(\delta r,\theta,z)$ centered on the track. These relate to $(x,y,z)$ by
\begin{equation}
\begin{aligned}
    x &= A \cos(z/f) + \delta r \cos \theta  \\
    y &= A \sin(z/f) + \delta r \sin \theta \\
    z &= z
\end{aligned}.
\end{equation}
The new field space metric becomes
\begin{align}
\mathcal{G}_{IJ} &= \begin{pmatrix}
1 & 0 & -\frac{A}{f} \sin(z/f-\theta) \\
0 & \delta r^2 &  \frac{A}{f} \delta r \cos(z/f-\theta) \\
-\frac{A}{f} \sin(z/f-\theta) & \frac{A}{f} \delta r  \cos(z/f-\theta) & 1+A^2/f^2
\end{pmatrix}
\end{align}
and the potential takes the simpler form $V = \Lambda^4 \left( e^{z/R} + \Delta \left(1-\exp\left[-\frac{\delta r^2}{2\sigma^2}\right] \right) \right)$. In these coordinates $\epsilon_V$ takes the form
\begin{align}
    \epsilon_V = \frac{\Delta ^2 \delta r^2 R^2+\sigma ^4 e^{\frac{\delta r^2}{\sigma ^2}+\frac{2 z}{R}}}{2 R^2 \sigma ^4 \left(\Delta -e^{\frac{\delta r^2}{2 \sigma ^2}} \left(\Delta +e^{z/R}\right)\right)^2}.
\end{align}
For later convenience we also define the Cartesian offsets from the center of the track $\delta x = x - A \cos(z/f)$ and $\delta y = y- A\sin(z/f)$, such that $\delta r^2 = \delta x^2 + \delta y^2$.

This potential can satisfy the refined dS conjecture in a wide inflationary region. At constant $z$ and far away from the track or along its center ($\delta r \rightarrow 0,\infty$), $\epsilon_V \rightarrow 1/(2 R^2)$, so we can fulfill the conjecture in this limit with the choice of a small enough $R$. The walls of the track are always steeper than its center, so the entire track fulfills the conjecture if the center does. However, away from the track in the $z\rightarrow -\infty$ limit, $\epsilon_V \rightarrow 0$, so this region of the potential does not fulfill the conjecture. Because this region is distant from the inflationary region ($\epsilon_V = 10^{-3}$ at $\Delta \phi \sim 7.3 \Mpl$ from the start of inflation using the parameters in Figure \ref{fig:helixTanhBackground}) and the swampland distance conjecture tells us that we should not trust effective field theories in asymptotic field space anyway, we do not consider its presence to diminish our argument.

 \subsection{Background dynamics}
The background equations of motion are
\begin{equation}
\begin{aligned}
\centering
&H^2=\frac{V}{3-\epsilon_H} \\ \\
&{\delta r}^{\prime \prime} + (3-\epsilon_H) {\delta r}^\prime - \delta r\,{\theta^\prime}^2 + \frac{A}{f^2} \cos(z/f-\theta) {z^\prime}^2 + \\
&+\frac{\Lambda^4 }{H^2} \left(\frac{\Delta \delta r}{\sigma^2} \,e^{-\delta r^2/(2 \sigma ^2)} \left(1+\frac{A^2}{f^2}-\frac{A^2}{f^2} \cos^2\left(z/f-\theta \right)\right)+\frac{A e^{z/R}}{f R} \sin \left(z/f-\theta \right)\right) = 0 \\ \\
&\theta^{\prime \prime} + (3-\epsilon_H)\theta^\prime + 2\frac{\delta r^\prime \theta^\prime}{\delta r} - \frac{A}{f^2\delta r} \sin\left(2z/f-2\theta\right){z^\prime}^2 +\\ &+\frac{\Lambda^4 A}{2 f^2 H^2} \left( -\frac{A \Delta  e^{-\delta r^2/(2 \sigma^2)} \sin \left(2 z/f-2 \theta\right)}{\sigma^2}-\frac{2 f e^{z/R} \cos \left(z/f-\theta \right)}{R \delta r} \right) = 0 \\ \\
&z^{\prime \prime}+ (3-\epsilon_H) z^\prime + \frac{\Lambda^4}{H^2} \left(\frac{A \Delta  \delta r}{f \sigma ^2} e^{-\delta r^2/(2 \sigma ^2)} \sin \left(z/f-\theta\right)+\frac{1}{R}e^{z/R}\right)= 0
\end{aligned}
\label{equ:helixeom}
\end{equation}
where primes denote e-fold derivatives $\partial_t \equiv H \partial_N$.
Note that the background evolution depends only on the combination $\Lambda^4/H^2$, which is independent of $\Lambda$. However, the same is not true for the perturbations. We will later exploit this to set the amplitude of the scalar perturbations without affecting the background evolution.

There is a steady-state solution with the fields approximately centered in the helical track, which we give in appendix \ref{sec:helix_soln}. We term this solution ``steady-state" because all of its slow-roll parameters are constants in time. It is analytically tractable, and gives high-slope inflation ($\epsilon_V \gg \epsilon_H$) in a wide region of parameter space. Numerically, small perturbations ($\Delta \delta r \lesssim \sigma /4$) around the trajectory converge to the steady-state solution. As we show in the appendix, while classically viable, this solution generates an observationally-excluded tensor power so we will not study it here.

\begin{figure}[t]
\centering
\includegraphics[width=\textwidth]{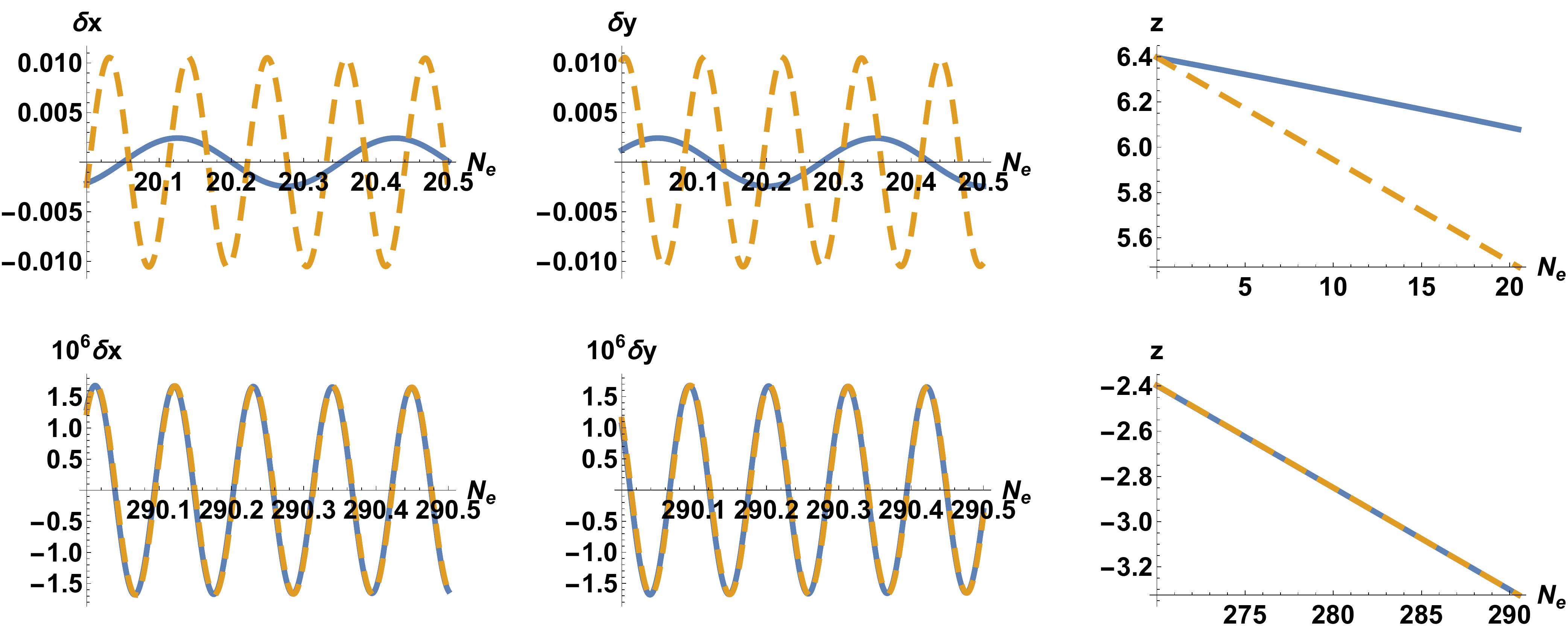}
    \caption{A comparison of metastable (blue) and steady-state (dashed, yellow) dynamics in the helix potential \eqref{eq:helixpotential}, without the inflation-ending modification \eqref{eq:Dz}. In the top row, we plot $(\delta x,\delta y,z)$ in units of $\Mpl$ at an early time, when the metastable solution has a smaller $z^\prime$, and smaller orbit around the helix's center in the $x,y$-plane. In the bottom row, the metastable solution has converged to a steady-state solution (with a different initial $z$ than the metastable solution) by $N_e \simeq 290$.}
\label{fig:metaVsSteady}
\end{figure}

\begin{figure}[t]
\centering
\includegraphics[width=0.8\textwidth]{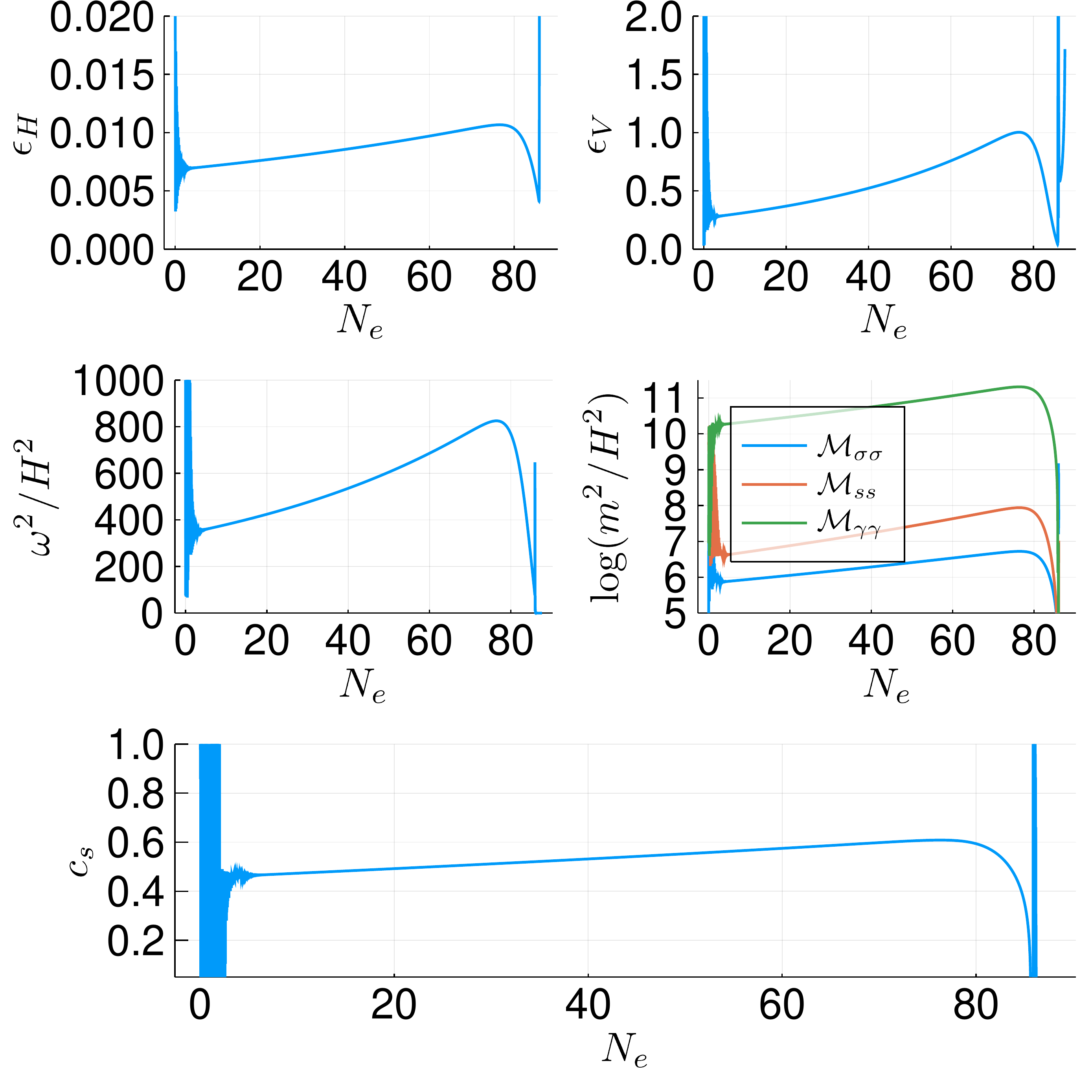}
\caption{Background evolution for the helix potential (\ref{eq:helixpotential}), with inflation-ending modification (\ref{eq:Dz}). Parameter values used were $R=1.07 \Mpl$, $\Delta_0=8.44$, $A = 3.47\times10^{-3}\Mpl$, $f=7.85\times10^{-4}$, $\sigma=8.9\times10^{-3} \Mpl$, $f_t=0.1\Mpl$. The initial value of $z$ and $z_\mathrm{end}$ were chosen to give $87.6$ e-folds. The excursion covered is large, but not asymptotically so $\Delta \phi \sim 1.5 \Mpl$ ($3.1 \Mpl$) in the unmodified (modified) potential. The difference in excursion, compared to the unmodified case, occurs during the last 0.1 e-folds.}
\label{fig:helixTanhBackground}
\end{figure}

Some perturbations around the steady-state solution only converge to it at late times -- we term these ``metastable" solutions. As we show in subsection \ref{sec:helixpert} below, this model's Planck-compatible regions of parameter space correspond to this class of solutions. We lack an analytic description of these metastable dynamics\footnote{Many of our parameter selections were found with a differential evolution optimizer from the \texttt{BlackBoxOptim.jl} package \cite{Feldt2018}, applied to our Julia-language implementation of the transport method \cite{Dias:2015rca}.}, which we compare to the steady-state dynamics in Figure \ref{fig:metaVsSteady}. In brief, the metastable solution falls much more slowly in the $z$ direction, with $z^\prime_\textrm{metastable} \sim z^\prime_\textrm{steady-state}/3$ initially. Due to the decreased velocity down the helix, the fields also stay closer to its center, with $\delta x_\textrm{metastable}$ and $\delta y_\textrm{metastable}$ smaller than their steady-state counterparts. We present the slow-roll parameters of one realization of the metastable solution in Figure \ref{fig:helixTanhBackground}.
These parameters give $\epsilon_V \sim 0.5$ outside the track, and a turning rate $\omega^2 / H^2\sim 10^4$.

The steady-state solution has a constant $\epsilon_H$ as is common to exponential inflation, so inflation does not end. The metastable solution in the regime of interest has $\epsilon_H < \epsilon_{H \textrm{ steady-state}}$, so it also cannot end inflation. However a small modification to the potential can end inflation without affecting the background behavior during the moment of horizon-crossing, and leave the perturbations invariant. We take the depth of the track to vary along the motion, so that it decreases as
\begin{align}
    \Delta(z) = \Delta_0 \tanh\left( \frac{z-z_\mathrm{end}}{f_t} \right)
\label{eq:Dz}
\end{align}
where $\Delta(z_0) \approx \Delta_0$, and $\epsilon_H \rightarrow 1$ occurs approximately when $z\rightarrow z_\mathrm{end}$.
\FloatBarrier
\subsection{Perturbations}
\label{sec:helixpert}
As can be seen in Figure \ref{fig:helixTanhBackground}, the two entropic masses are both larger than the adiabatic mass. 
 With heavy isocurvature masses, multi-field inflationary scenarios with high turning rates produce perturbations similarly to single-field models with a reduced speed of sound $c_s$.
 This has been rigorously derived for two-field scenarios in {\cite{Achucarro:2012sm, Achucarro:2012yr,Hetz:2016ics}} and for three-field scenarios in \cite{Cespedes:2013rda}. This effective single-field theory becomes more accurate as the gap between the adiabatic and entropic masses grows.
Although in our case the masses are not parametrically separated, we can study the single-field effective theory from integrating them out, knowing that it is subject to $\cM_{\sigma\sigma} / \cM_\textrm{entropic}$ corrections. 

In our notation, the effective speed of sound is
\begin{equation}
\begin{aligned} \label{eq:cs}
\frac{1}{c_s^2} &= 1 + \frac{4 \omega^2 (V_{\gamma \gamma} - |\dot{\gamma}|^2 )}{\det{M}} \\
M&\equiv \begin{pmatrix}
V_{ss} - \omega^2 - |\dot{\gamma}|^2  & V_{s \gamma} \\
V_{s \gamma} & V_{\gamma \gamma} - |\dot{\gamma}|^2 \\
\end{pmatrix}
\end{aligned}
\end{equation}
where $s^I$ and $\gamma^I$ are normal and binormal unit vectors to the trajectory as in (\ref{eq:defsg}). For the fields $x,y,z$ in flat space, $\hat{\gamma} \equiv \hat{\sigma}\times \hat{s}$. 

The perturbations of single-field models with reduced speed of sound are well studied \cite{Achucarro:2012sm,Achucarro:2012yr,Burgess:2012dz}.
The spectral tilt in such models is
\begin{align}
n_s - 1 \simeq -2 \epsilon_H - \eta_H - \kappa
\label{eq:singleFieldns}
\end{align}
where $\kappa \equiv c_s^\prime/c_s$. 
Similarly the tensor-to-scalar ratio is given by
\begin{align} \label{eq:singleFieldr}
r=16\epsilon_H c_s
\end{align}
which is suppressed in the small $c_s$ limit. The equilateral non-gaussianity is inversely proportional to $c_s$, however, so the sound speed cannot be made arbitrarily small and agree with observations.
\begin{align}
f^{(\rm{eq})}_{NL}= \frac{125}{108} \frac{\epsilon_H}{c_s^2}+ \frac{5}{81} \frac{c^2_s}{2} \left(1-\frac{1}{c^2_s} \right)^2+\frac{35}{108} \left(1-\frac{1}{c^2_s} \right)
\label{eq:singleFieldfnl}
\end{align}

\begin{figure}[h]
\centering
\includegraphics[width=\textwidth]{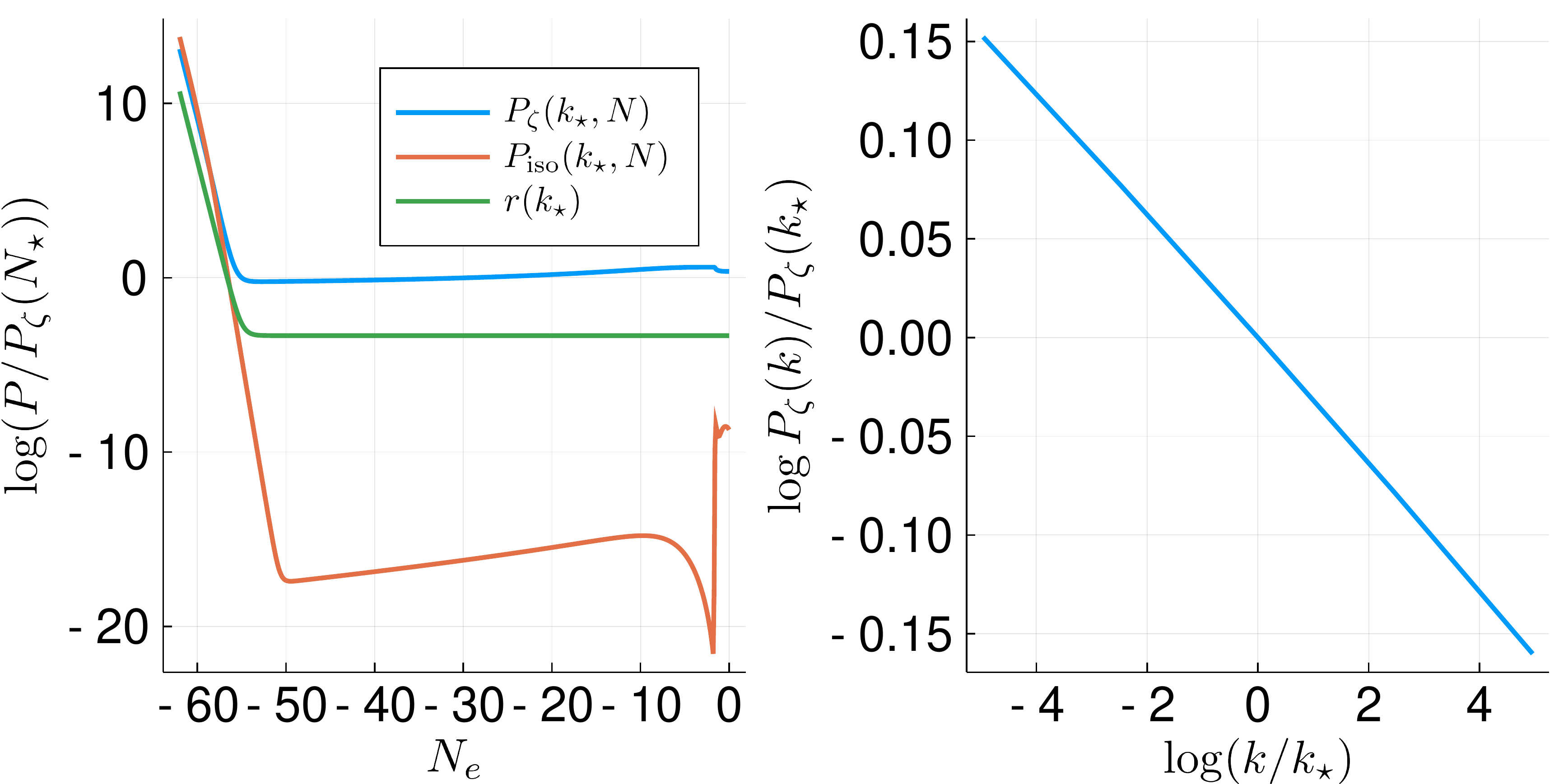}
\caption{Powerspectra of scalar and tensor modes during the background dynamics of Figure \ref{fig:helixTanhBackground}, evolved with the transport method. The plot begins when we impose Bunch-Davies initial conditions for the modes, and ends at the end of inflation. Horizon exit occurs at $N_\star=-55$ for the pivot scale $k_\star=0.002 \textrm{ Mpc}^{-1}$. This simulation has tensor-to-scalar ratio $r=0.036$ and the spectral index $n_s=0.9685$. At the end of inflation, isocurvature power is suppressed by a factor of $\sim 10^4$.}
\label{fig:helixTanhPowerspectra}
\end{figure}

In our steady-state solution, $\eta_H \approx \kappa \approx 0$, and $c_s \gtrsim 0.8$. This solution, then, influences $n_s$ only by the effects of $\epsilon_H$, and a large $\epsilon_H$ is needed for $n_s$ to be Planck-compatible. Unfortunately, this also raises the expected tensor power, $r$, observationally excluding this solution.
Our metastable solution, however, has a much smaller $c_s$ and cannot be excluded by the same reasoning. $\eta_H$ and $\kappa$, while small, are not negligible.
This argument was verified with a full transport method evolution of the perturbations, which is equivalent to tree-level in the in-in formalism (see appendix \ref{sec:transport} for a brief overview of the method). The powerspectra corresponding to the background evolution in figure \ref{fig:helixTanhBackground} are shown in figure \ref{fig:helixTanhPowerspectra}.

We emphasize these models have low isocurvature ($r_\textrm{iso} \equiv P_\mathrm{iso}(k_\star,N_\textrm{end})/P_\zeta(k_\star,N_\textrm{end}) \sim 10^{-4}$ ) and a featureless adiabatic powerspectrum. The computed values of $n_s$ and $r$ lie within the $2\sigma$ Planck ellipse in the $n_s$-$r$ plane.
Tensor modes are speed-of-sound supressed, with $c_s \sim 0.5$.

We did not perform any numerical analysis of the bispectrum of perturbations, but the effective single-field result can give an estimate. 
For the horizon-exit value of an $\epsilon_H\sim 0.006$, a sound speed of $0.1 \lesssim c_s$ is consistent with the $1\sigma$ value from Planck 2018 \cite{Akrami:2018odb,Akrami:2019izv}. With the speed of sound in figure \ref{fig:helixTanhPowerspectra} we estimate $f^{(\rm{eq})}_{NL}\simeq -18$, well within the $1\sigma$ bound.

The EFT results for $n_s$ and $r$ do not agree exactly with their transport-method counterparts: $n_s|_\mathrm{EFT} - n_s|_\mathrm{transport} \sim 0.005$ and $r|_\mathrm{EFT} = 0.0676 \sim 1.87 r|_\mathrm{transport}$. The EFT is not in its full regime of validity, due to the relatively small mass gap between $\cM_{\sigma \sigma}$ and $\cM_{ss}$. Nonetheless, we expect a full numerical calculation of $f^{(\rm{eq})}_{NL}$ to be Planck-compatible, given that we would need almost an effective $c_s$ a factor of 5 lower to reach $1\sigma$ tension with the Planck result. It would be interesting to explore the full transport-method evolution and shape of this model's bispectrum.
In addition to equilateral non-gaussianity, $c_s \neq 1$ trajectories have also been proposed to generically source large non-gaussianities in folded configurations \cite{Tolley:2009fg,Achucarro:2010da}.

The model presented here is the first flat field space high-turning rate inflationary solution consistent with the dS-conjecture -- it also has Planck-compatible regions of parameter space. The predicted tensor power is relatively high, and within the range of upcoming experiments, e.g. LiteBIRD \cite{Hazumi:2019lys}. Effective reduced speed-of-sound models virtually guarantee either relatively large tensor modes ($r \gtrsim 10^{-3}$) or large equilateral non-gaussianity, and increased bounds on both would tightly constrain this type of track-like construction.

\section{Superpotential model}
\label{sec:super}
In this section we study, and then modify, an analytically simple model with negative field-space curvature studied by Chen et al. \cite{Chen:2018uul} in the context of primordial non-gaussianities. This is a particular case of a larger family of potentials analyzed in \cite{Achucarro:2019pux, Christodoulidis:2019mkj} that is conducive to analytical results. We begin by highlighting the model's desirable classical behavior: it can achieve a sufficiently high number of e-folds of inflation with distinctly non-geodesic motion, while globally satisfying the refined de Sitter conjecture (\ref{eq:revswamp}). We then discuss the challenges of making the quantum fluctuations' phenomenology sound. The single-field EFT used in section \ref{sec:helix} breaks down, so analysis of the perturbations must be entirely two-field. When combined with the negative field-space curvature, a high rate of turning is known to yield an exponential growth of the perturbations before horizon exit. This is not fatal, but bounds the turning rate from above to prevent the scale of inflation from dipping below the nucleosynthesis scale. This section shows that it is possible to construct a model that satisfies all these constraints.

The model has two fields, $\phi^I = \{X,Y\}$, with a  hyperbolic metric:
\begin{align}\label{cp_metric}
\cG_{IJ} = \begin{pmatrix}
e^{2 Y/R_0} & 0 \\
0 & 1 \\
\end{pmatrix}
\end{align}
with non-vanishing Christoffel symbols and Riemann tensor components
\begin{align}\label{cp_symbols}
\begin{aligned}
\Gamma^X_{XY} &= \Gamma^X_{YX} = \frac{1}{R_0} \\
\Gamma^Y_{XX} &= -\frac{1}{R_0}e^{2Y/R_0}
\end{aligned}
&&
\begin{aligned}
R^Y_{XXY} = -R^Y_{XYX} &= \frac{1}{R_0^2}e^{2Y/R_0} \\
R^X_{YYX} = -R^X_{YXY} &= \frac{1}{R_0^2}.
\end{aligned}
\end{align}
The potential is built from a ``superpotential," $W = W(X)$:
\begin{align}\label{cp_potential}
V(X,Y) = 3 W^2 - 2 \cG^{IJ} W_{,I} W_{,J}.
\end{align}
Note that due to the minus sign and the dependence on real-valued fields, $W$ is not a true superpotential; we use the term simply for convenience. This form of the potential\footnote{Note that models with a similar field space metric and different potentials have been presented in \cite{Achucarro:2016fby,Cremonini:2010sv}} can be realized in two ways: by demanding that inflation occurs along an isometry direction of the metric \cite{Achucarro:2019pux}, or by enforcing $\dot{Y}=0$ and a potential of the form $V(X,Y) = h(X) + f(X)g(Y)$; see Appendix \ref{SuperDerivation} for a derivation. The dynamics are given by:
\begin{align}
\dot{\phi}^I &= -2\cG^{IJ}W_{,J} = \left( -2e^{-2Y/R_0} W_X, 0 \right) \label{superEoM} \\
H &= W \label{superHubble} \\
\omega^I &= \left( 0, \frac{2}{R_0}e^{-Y/R_0}W_{,X} \right) \label{superOmega} \\
\epsilon_H &= \frac{R_0^2}{2}\frac{\omega^2}{H^2}.\label{superEpsilonH}
\end{align}

In consideration of the de Sitter conjecture, we can constrain the geometric scale $R_0$ by imposing high-turning, slow-roll inflation independent of the form of the superpotential: choosing $\omega/H \gtrsim 10^1$ and $\epsilon_H \lesssim 10^{-2}$ fixes $R_0 \lesssim 10^{-2}$.

An exponential superpotential,
\begin{align}
	W = A e^{X/R_1},
\end{align}
can easily meet $\epsilon_H \ll 1$, $\epsilon_V \gtrsim 1$, and $\omega/H \gg 1$ along the trajectory for all time. We find the following analytic results:
\begin{align}
Y(N) &= Y_0, \quad X(N) \equiv X_N = X_0 - \frac{2}{R_1}e^{-2Y_0/R_0}N \label{expSuperTrajectory} \\
N(t) &= \frac{R_1}{2}e^{2Y_0/R_0}\log\left[\frac{2A}{R_1^2}e^{X_0/R_1+2Y_0/R_1}t + 1\right] \\
\frac{\omega}{H} &= \frac{2}{R_0 R_1}e^{-Y_0/R_0} \\
\epsilon_H &= \frac{2}{R_1^2}e^{-2Y_0/R_0} \\
\epsilon_V &= \frac{2}{R_1^2}e^{-2Y/R_0} + \frac{8}{R_0^2}\frac{1}{\left( 3R_1^2e^{2Y/R_0} - 2 \right)^2}.\label{expSuperEpsilonV}
\end{align}
Here, $N$ is the number of e-folds elapsed since the start of inflation, and $t$ is cosmic time. We note that the superpotential scale $R_1$ cannot be chosen independently of $Y_0$ while maintaining slow-roll inflation. Using the above constraint on $R_0$, we find $e^{-Y_0/R_0}/R_1 = \sqrt{\epsilon_H/2} \lesssim 10^{-1}$. Hence, there exists a one-dimensional family of values for $R_1$ and $Y_0$ with the desired inflationary behavior. We further emphasize that $\epsilon_V$ parametrizes potential gradients throughout the entire field space, whereas the dynamical expressions above pertain to a particular inflationary trajectory.

We observe that the inflationary trajectory (\ref{expSuperTrajectory}) proceeds in the negative $X$ direction at a fixed value of $Y$. The ``turning" of this path can be seen by comparing against geodesics of this field space, which take the form:
\begin{align}\label{cp_geodesic}
\tilde{Y}(X) = R_0\log\left[ \frac{R_0}{\sqrt{C-X}\sqrt{K-C+X}} \right].
\end{align}
A derivation is presented in Appendix \ref{geoDerivation}. The constants $K$ and $C$ may be chosen such that the geodesic passes through any two points $(X_1,Y_1)$ and $(X_2,Y_2)$ such that $X_1 \neq X_2$; their values are given in (\ref{geoK}) and (\ref{geoC}). Evidently, the trajectory (\ref{expSuperTrajectory}) is strongly non-geodesic, with rapid turning for appropriately chosen parameters $R_0$, $R_1$, and $Y_0$.

\begin{figure}
	\centering
	\includegraphics[width=0.7\textwidth]{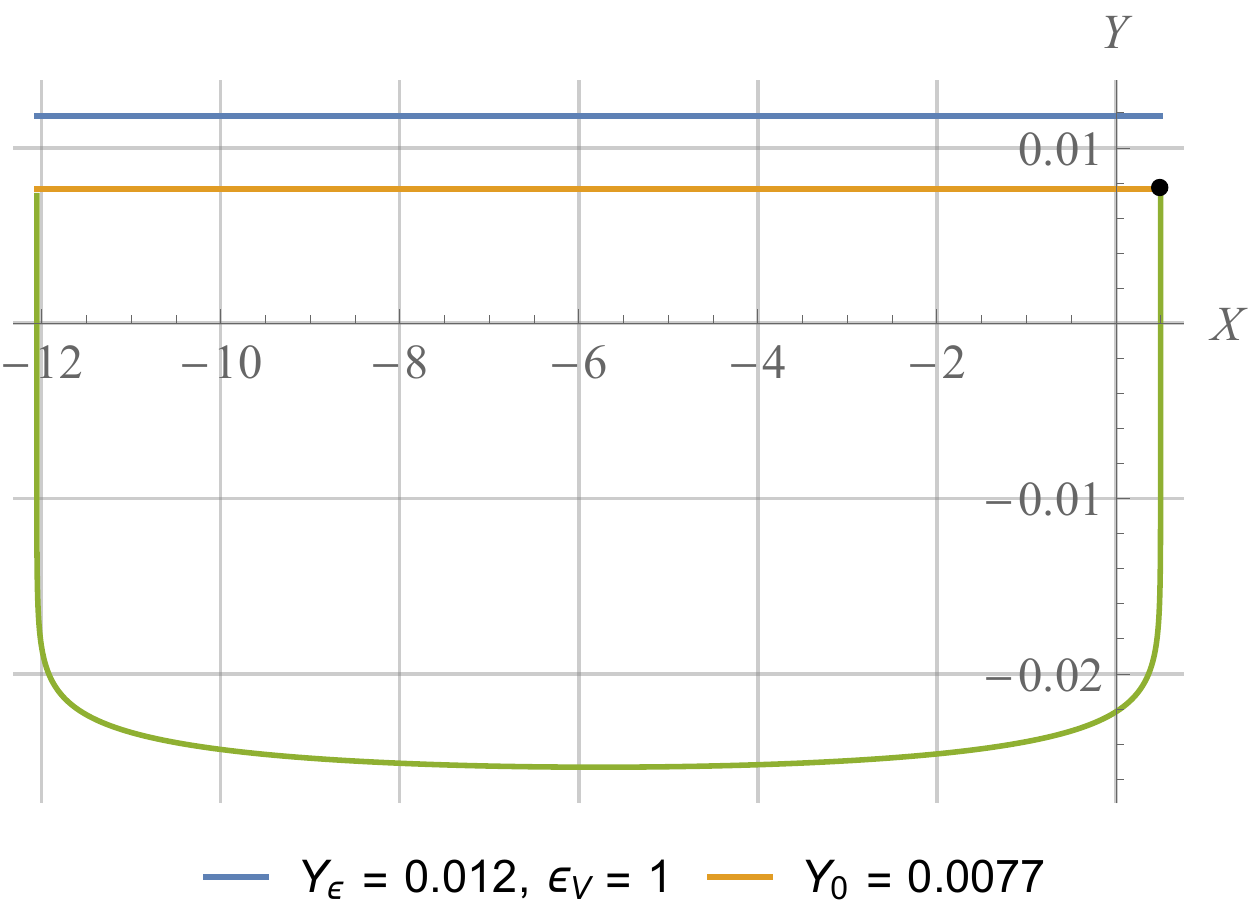}
	\caption{A trajectory of the form (\ref{expSuperTrajectory}) with $N=300$ e-folds of inflation with $X_0=0.5$, $X_N=-12.05$, $Y_0 = 0.0077$, $R_0=0.0034$, and $R_1=0.5$. The geodesic connecting $(X_0,Y_0)$ to $(X_N,Y_0)$ extends below the trajectory and yields a field excursion of 0.071 $\Mpl$. The line at $Y_\epsilon=0.012$ corresponding to $\epsilon_V=1$ is highlighted as well, with $\epsilon_V > 1$ everywhere below this line.}
	\label{fig:ExpSuperTrajectory}
\end{figure}

An important feature of this class of trajectories is that field excursions are easily made sub-Planckian due to the small value of $R_0$. This ensures that the effective field theory with scalar potential (\ref{cp_potential}) does not break down over the course of inflation \cite{Obied:2018sgi, Ooguri:2018wrx}. The excursion is defined as the geodesic distance between two points $(X_i,Y_i)$ and $(X_f,Y_f)$ and is given by (\ref{App_geolength}). We consider a trajectory from the initial point $(X_i,Y_i) = (X_0,Y_0)$ up until the point corresponding to $N$ e-folds of inflation $(X_f,Y_f) = (X_N,Y_0)$. Choosing a geodesic that passes through these points, the expressions for $K$ and $C$ simplify:
\begin{align}
K &= \sqrt{(X_N-X_0)^2 + 4Q} \\
C &= \frac{X_N + X_0 + K}{2},
\end{align}
where $Q = R_0^2 e^{-2Y_0/R_0}$. The geodesic distance (\ref{App_geolength}) reduces to
\begin{align}
S = \frac{R_0}{2}\log\left[ \left(\frac{X_0 - X_N + K}{X_0 - X_N - K}\right)^2 \right].
\end{align}
The small geometric scale $R_0$, required to have high-turning inflation, strongly suppresses the distance for many possible values of $X_0$ and $X_N$, which ensures that the potential is valid throughout inflation. A sample trajectory with rapid turning and sub-Planckian field excursion is displayed in Figure (\ref{fig:ExpSuperTrajectory}) with the corresponding geodesic connecting $X_0$ and $X_N$.

\begin{figure}
	\centering
	\includegraphics[width=0.8\textwidth]{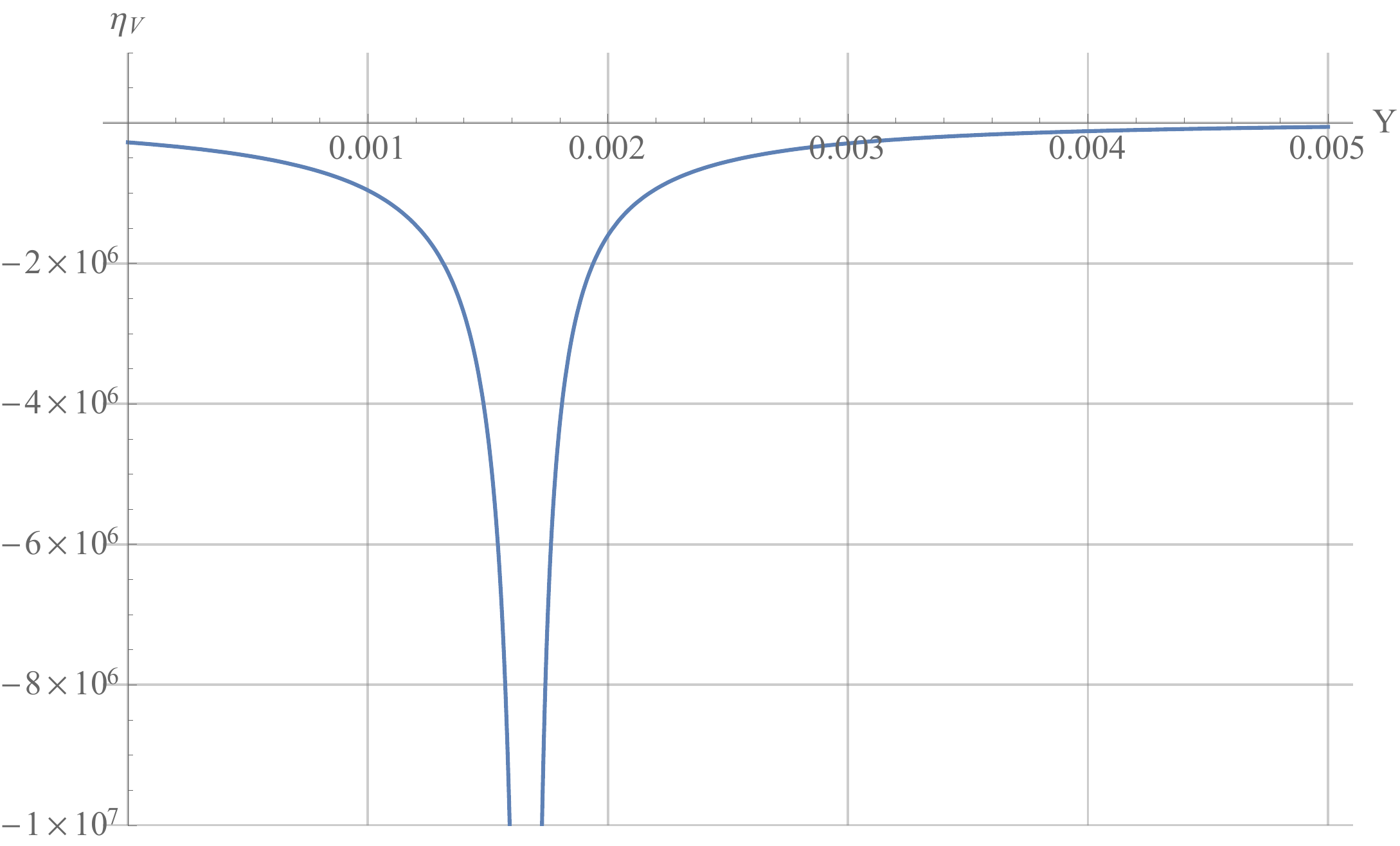}
	\caption{The minimum eigenvalue of the covariant Hessian matrix for $R_0=0.0034$ and $R_1=0.5$. This quantity is independent of $X$. The asymptotic behavior arises from the potential $V$ vanishing at $Y=0.00167$ and becoming negative below this value. For large positive and negative $Y$, $\eta_V$ asymptotes to approximately $-3$ and $-20$ respectively, thus satisfying the refined de Sitter swampland conjecture.}
	\label{fig:ExpSuperEtaV}
\end{figure}

We note that $\epsilon_V$ vanishes for $Y \gg R_0$. Although this conflicts with the gradient swampland conjecture, we find that the refined de Sitter conjecture \cite{Ooguri:2018wrx} still holds. In particular, this model satisfies the right half of (\ref{eq:revswamp}): $\eta_V$ is globally negative. Since the superpotential is chosen to be an exponential, $\eta_V$ is independent of $X$ for this model. Figure (\ref{fig:ExpSuperEtaV}) displays $\eta_V$ as a function of $Y$, with asymptotic values that are $\mathcal{O}(-1)$. Therefore, this potential indeed satisfies the refined de Sitter conjecture globally, in spite of $\epsilon_V$ vanishing for $Y$ sufficiently large.

Furthermore, we observe that there are values of $Y$ for which neither conjecture regarding the potential need be satisfied, namely those at which the potential becomes negative:
\begin{align}
Y < \frac{R_0}{2}\log\left[ \frac{2}{3(R_1)^2} \right].
\end{align}
The classical solution does not yield inflation in this region, since this corresponds to $\epsilon_H > 3$ along the trajectory. Hence, initial conditions must be chosen such that this region is avoided.

The primary drawback to this model lies in $\eta_H$ being identically $0$, so inflation does not end\footnote{In the sense of \cite{Christodoulidis:2019jsx}, background motion with constant slow-roll parameters can be seen as the critical point of a dynamical system.}. In order to terminate inflation, we seek an alternate superpotential such that $\epsilon_H$ crosses unity from below. In particular, we choose a function that preserves the exponential behavior at asymptotic values of $X$. Consider:
\begin{align}\label{superExpTanh}
W(X) = A e^{X/R_1}\left[\tanh\left(\frac{X}{R_2}\right) + 1\right].
\end{align}
This modification preserves the positivity of the superpotential, and hence the Hubble parameter. From (\ref{superEoM}), the equations of motion are:
\begin{align}
\dot{X} = -2A e^{-2Y_0/R_0}e^{X/R_1}\left[\frac{\tanh(X/R_2)}{R_1} + \frac{\text{sech}^2(X/R_2)}{R_2}\right], \quad \dot{Y} = 0.
\end{align}
The slow-roll parameter for this model is of the form
\begin{align}
\epsilon_H = \frac{2 e^{-\frac{2 Y}{R_0}} \left[2 R_1 + R_2 \left(e^{\frac{2 X}{R_2}} + 1\right)\right]^2}{R_1^2 R_2^2 \left(e^{\frac{2 X}{R_2}} + 1\right)^2},
\end{align}
which increases monotonically as $X$ decreases from its initial value. This ensures that inflation terminates after a finite number of e-folds.

\begin{figure}
	\centering
	\includegraphics[width=0.8\textwidth]{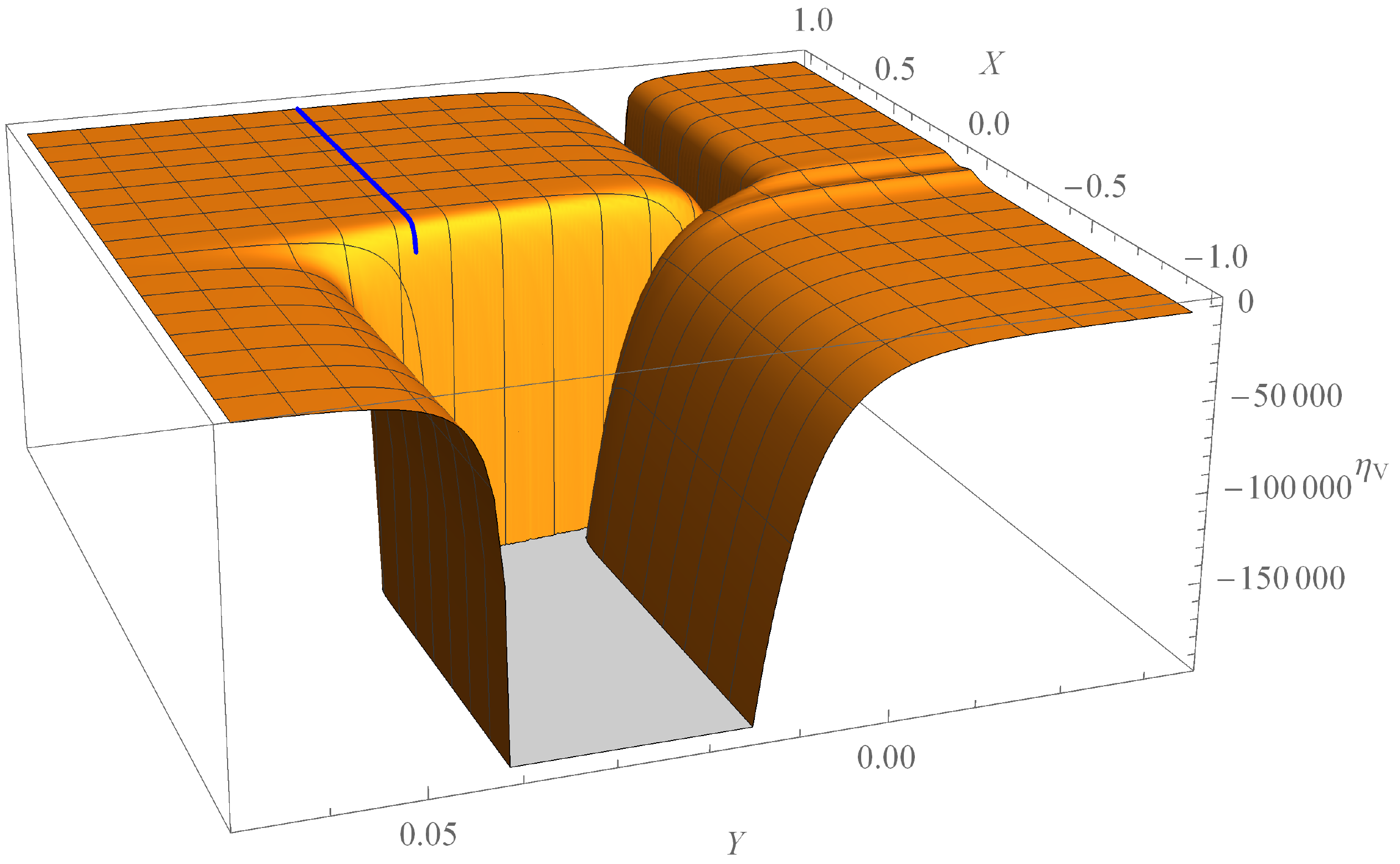}
	\caption{The minimum eigenvalue of the covariant Hessian matrix in the vicinity of the trajectory for (\ref{superExpTanh}) with $R_0=0.011$, $R_1=1.82$, and $R_2=0.05$. For large positive $Y$, $\eta_V$ asymptotes to $-3.00$. For large negative $Y$ and large positive $X$, $\eta_V$ asymptotes to $-1.51$, becoming even more negative for large negative values of $X$. For negative values of $Y$ in the vicinity of $X=0.15$, $\eta_V$ has a local maximum that is positive, but $\epsilon_V > 1$ in this region. Hence, the refined de Sitter swampland conjecture is satisfied. The value of $\eta_V$ along the trajectory is highlighted in blue. The trajectory is chosen to begin at $X_0 = 1$ along a constant value of $Y_0 = 0.034$. The turning rate and slow-roll parameter are approximately constant at $\omega/H = 4.84$ and $\epsilon_H = 0.0015$, respectively, until the very end of inflation, when $\epsilon_H$ quickly rises and crosses unity at $X_N = 0.016$. This trajectory yields $N = 328$ e-folds of inflation and a field excursion of 0.17 $\Mpl$.}
	\label{fig:ExpTanhEtaV}
\end{figure}

Unlike (\ref{expSuperEpsilonV}), the value of $\epsilon_V$ for this superpotential depends on both fields $X$ and $Y$, although it still does not globally remain $\mathcal{O}(1)$ or larger. However, parameters of the model can be chosen such that $\eta_V$ is bounded from above by $\mathcal{O}(-1)$ values everywhere, except in a one-dimensional region near $X = 0$ where the $\tanh$ factor dominates; see Figure (\ref{fig:ExpTanhEtaV}). Fortunately, the parameter space allows for $\epsilon_V \gtrsim 1$ in this region, so long as the turning rate on the trajectory is sufficiently large. Therefore, the refined de Sitter conjecture remains satisfied.

The quantum perturbations in this model are well studied, since it is a subclass of orbital inflation \cite{Achucarro:2019pux} and of the broader category of models studied in \cite{Christodoulidis:2019mkj}. A notable feature of this class of one-field superpotentials is the simplicity of the first order equations of motion, which allows for the mass-squared matrix (\ref{eq:massMatrix}) to be easily computed. The adiabatic component is of the form:
\begin{align}\label{eq:massMatrixAdiabatic}
	\cM_{\sigma\sigma} = \omega^2 + 6H^2\epsilon_H - \frac{3}{2}H^2\eta_H + \frac{5}{2}H^2\epsilon_H\eta_H - \frac{1}{4}H^2\eta_H^2 - 2H^2\epsilon_H^2 - \frac{1}{2}H\dot{\eta}_H.
\end{align}
Note that for high-turning inflation, $\cM_{\sigma\sigma} \simeq \omega^2 + \mathcal{O}\left(\epsilon_H,\eta_H \right)$. The entropic component has the form
\begin{align}\label{eq:massMatrixSuper}
\mathcal{M}_{ss} = -\frac{12}{R_0^2}e^{-2Y_0/R_0}W_X^2 = -3\omega^2.
\end{align}
Hence on subhorizon scales, the effective entropic mass $\mu_{s,sub}^2 = \mathcal{M}_{ss} - \omega^2$ is negative and large in magnitude, due to the nontrivial turning rate\footnote{This precise value of the entropic mass violates the criteria of validity for the usual single-field EFT, see appendix A of \cite{Garcia-Saenz:2018vqf}. The speed of sound ($c_s^{-2}=1+4\omega^2/\mu_{s,\mathrm{sub}}^2$) diverges. For this reason, our analysis in this section is strictly two-field.}$^,$\footnote{These masses' dependence on $\omega$ agrees with the rapid-turn inflationary attractor \cite{Bjorkmo:2019fls}.}. From (\ref{eq:entropicEomSubSuper}) and (\ref{eq:adiabaticEOM}), we see that this sources an exponential growth of both the adiabatic and entropic power for modes with $k^2/(aH)^2 < 4\omega^2/H^2$. This is a manifestation of the tachyonic instability of modes with a large and negative $\mu_{s,sub}^2/H^2$ discussed in \cite{Fumagalli:2019noh}.

On superhorizon scales our model has an exactly massless entropic perturbation, as shown in \cite{Achucarro:2019pux} and \cite{Christodoulidis:2019mkj}. This is a consequence of a flat direction in the effective potential, whose gradient is
\begin{align}
V^{\alpha}_\mathrm{eff} = V^{\alpha} + 2\epsilon_H H^2 \Gamma_{\sigma \sigma}^\alpha,
\end{align}
where $\alpha$ indexes non-adiabatic directions. In this case, $V_\mathrm{eff}^y$ is identically zero.

After horizon exit, entropic perturbations will freeze, as seen from (\ref{eq:entropicEomSubSuper}). As long as $\omega/H\gtrsim 1$, the entropic modes will feed the growth of the adiabatic modes causing them to grow linearly with time. This feature reduces the ratio of the entropic power over the adiabatic power as inflation continues and $\omega/H$ remains large.

We now estimate the degree of exponential growth, $x$, of the power spectrum:
\begin{align}\label{powerExponentialGrowth}
P_\cR = \frac{H^2_\star}{8\pi^2 \epsilon_{H\star}} e^{2x}.
\end{align}
Following the notation of \cite{Bjorkmo:2019fls,Bjorkmo:2019qno}, we express the entropic component of the mass-squared matrix (\ref{eq:massMatrix}) as:
\begin{align}
	\cM_{ss} \equiv \xi\omega^2.
\end{align}
From (\ref{eq:massMatrixSuper}), we see that one-field superpotential models have $\xi = -3$. The equations of motion (\ref{eq:adiabaticEOM}) and (\ref{eq:entropicEOM}) can then be recast in terms of the curvature scalar $\cR_c$ as:\footnote{Rewriting the equations in this form also allows us to check the stability of the background trajectory by computing the Lyupanov exponents as in \cite{Bjorkmo:2019aev,Bjorkmo:2019fls}. To lowest order in slow-roll parameters, the exponents are $0,0,-3H,-3H$. All are zero or negative, ensuring stability.}
\begin{align}
	\ddot{\cR}_c + 3\left(H+\eta_H \right)\dot{\cR}_c + \frac{k^2}{a^2}\cR_c &= \frac{2\omega}{\sqrt{2\epsilon_H}}\left[\dot{Q}_s + \left(3-\epsilon_H \right)H Q_s \right] \label{eq:adiabaticEOM_Rc}\\
	\ddot{Q}_s + 3H\dot{Q}_s + \left[\frac{k^2}{a^2} + (\xi - 1)\omega^2 \right]Q_s &= -2\omega\sqrt{2\epsilon_H}\dot{\cR}_c. \label{eq:entropicEOM_Rc}
\end{align}
In order to find the growth parameter $x$, we employ the WKB method as used in \cite{Bjorkmo:2019qno}, without dropping the Hubble friction terms in (\ref{eq:adiabaticEOM_Rc}) and (\ref{eq:entropicEOM_Rc}). We note that the slow-roll suppressed terms in (\ref{eq:adiabaticEOM_Rc}) can be safely neglected in this computation. Anticipating an exponential amplification of both modes, we assume a solution of the form:
\begin{align}
	\cR_c = \cR_c^{(0)}e^{\lambda t}, \quad Q_s = Q_s^{(0)}e^{\lambda t}.
\end{align}
Inserting this ansatz into (\ref{eq:adiabaticEOM_Rc}) and (\ref{eq:entropicEOM_Rc}), enforcing $\lambda > 0$, and setting $\xi = -3$ for our model, we find:
\begin{align}
	\tilde{\lambda} \equiv \lambda / H = \frac{1}{2}\left[ -3 + \sqrt{9 - 4(\kappa^2 - 2\kappa\omega/H)} \right],
\end{align}
where $\kappa \equiv \frac{k}{aH}$. Note that a positive $\lambda$ requires $\kappa < 2\omega/H$; since $\kappa$ decays with the number of e-folds elapsed, the exponential growth begins at $N = -\log\left(2\omega /H \right)$ before horizon crossing. The adiabatic mode then grows as:
\begin{align}
	\cR_c \sim \exp\left[ \int_{-\log\left(2\omega / H\right)}^{N} \left|\tilde{\lambda} \right| dN' \right].
\end{align}
As shown in \cite{Achucarro:2019pux}, we can solve (\ref{eq:adiabaticEOM_Rc}) and (\ref{eq:entropicEOM_Rc}) on superhorizon scales to find:
\begin{align}
	Q_s = \frac{H_\star}{2\pi}, \quad \left|\cR_c \right|^2_\mathrm{super} = \frac{N_\star^2 \omega_\star^2}{2\pi^2 \epsilon_{H\star}}
\end{align}
The total adiabatic power including sub- and superhorizon contributions is then:
\begin{align}
	P_\cR = \frac{H^2_\star}{8\pi^2 \epsilon_{H\star}}\exp\left[ 2\int_{-\log\left(2\omega/H \right)}^{\infty} \left|\tilde{\lambda} \right| dN' \right] \left(1 + \frac{4N_\star^2 \omega_\star^2}{H_\star^2} \right).
\end{align}
Comparing this against (\ref{powerExponentialGrowth}), we solve for the effective growth parameter at the end of inflation:
\begin{align}\label{xGrowthParameter}
	x = \int_{-\log\left(2\omega/H \right)}^{\infty} \left|\tilde{\lambda} \right| dN' + \frac{1}{2}\log\left(1 + \frac{4N_\star^2 \omega_\star^2}{H_\star^2} \right).
\end{align}

\begin{figure}[t]
\centering
\includegraphics[width=\textwidth]{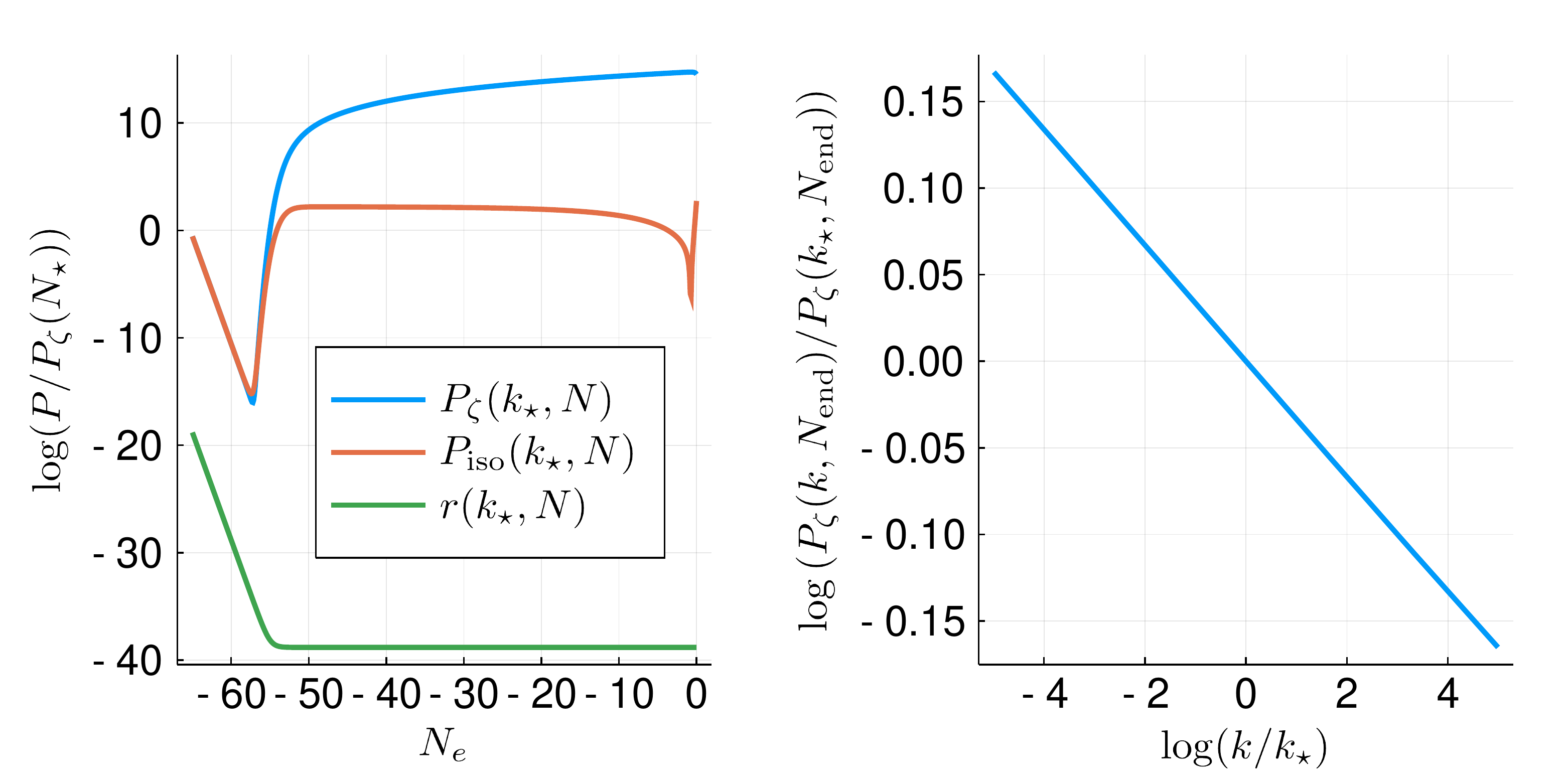}
\caption{Perturbative powerspectra for the pivot-scale mode in the superpotential model, with Bunch-Davies initial conditions numerically imposed well before any subhorizon growth at $10$ e-folds before horizon exit. The scalar powerspectra give $n_s = 0.966$ and $r_\textrm{iso}\sim10^{-6}$. Tensor modes are exponentially small, with both the sub- and superhorizon growth of the adiabatic mode giving $r\sim 10^{-17}$ by the end of inflation. We measure a growth parameter $x=15.371$, in good agreement with \eqref{xGrowthParameter}.}
\label{fig:superpotentialPowerspectra}
\end{figure}

Solving the integral numerically, this yields $x = 14.962$ for $\omega_\star/H_\star \sim 4.84$. Fitting the transport-method adiabatic power spectrum in Figure \ref{fig:superpotentialPowerspectra} yields $x = 15.371$. We observe that this agrees remarkably with the WKB calculation so long as the superhorizon contribution to the adiabatic power is also included. Including the Hubble friction terms in \eqref{eq:adiabaticEOM_Rc} was also important -- neglecting them predicts $x\sim 21$, a much higher growth. Comparing this against (\ref{powerExponentialGrowth}) and taking $\epsilon_H = 0.0015$ from Figure (\ref{fig:ExpTanhEtaV}), we find a value for the Hubble parameter:
\begin{align}
	H_* \sim 10^{-12} \Mpl \sim 10^6 \hspace{1ex}\text{GeV}.
\end{align}
Thus, the parameter space permits a mass scale of inflation compatible with nucleosynthesis bounds, $H_\text{min} \approx 4 \hspace{1ex}\text{MeV}$, so long as the turning rate is not excessively large.
Saturating the nucleosynthesis bound requires:
\begin{align}
	x = \frac{1}{2}\log\left( \frac{8\pi^2 \epsilon_{H\star}}{H_\text{min}^2} \right) \simeq 46 + \log\left( \frac{\epsilon_{H\star}}{0.0015} \right)^{1/2}.
\end{align}
For $\epsilon_{H\star} = 0.0015$, this sets an upper bound on the turning rate, $\omega \lesssim 15$.

In \cite{Fumagalli:2019noh}, the authors use the single-field EFT to estimate the growth of
flattened non-gaussianities as a function of the growth of the power spectrum
for models with an imaginary speed of sound. We are unable to use their result to reliably compute $f_\mathrm{NL}^\mathrm{flat}$, due to the invalidity of the EFT for our model.

In summary, superpotentials readily admit classical trajectories with $\epsilon_H \ll 1$, and either $\eta_V \lesssim -1$ or $\epsilon_V \gtrsim 1$ in all regions. The analysis of quantum perturbations shows that the entropic mode effective mass-squared of subhorizon modes, $\mu_{s,sub}^2 = \mathcal{M}_{ss} - \omega^2$, is large and negative, giving rise to an exponential growth of the perturbations in the subhorizon regime. In order to achieve a scale of inflation compatible with nucleosynthesis, the turning rate must be bounded from above. We emphasize that the parameter space admits trajectories with a desirable turning rate and phenomenology, while satisfying the refined de Sitter conjecture globally.

\section{Conclusions}
In this work we present two families of  multi-field potentials  with a high turning rate that support inflation while satisfying the de Sitter and the distance swampland conjectures. One family has a flat field space metric while the other's is negatively curved.  We analyze perturbations around the classical solutions and check their predictions against the current CMB experimental bounds. 

The flat field space model has three fields and a helix-like potential and has observationally consistent phenomenology. The predicted tensor power is relatively high, and within the range of upcoming experiments, e.g. LiteBIRD. Effectively, this model can be reduced to a single field model with reduced speed of sound. All such models virtually guarantee large tensor to scalar ratio or large equilateral non-gaussianity.  This potential does not globally satisfy the refined de Sitter conjectures (\ref{eq:revswamp}), but it does satisfy them around the inflationary trajectory in a region of at least $\mathcal{O}(\Mpl)$. We do not know of any UV-complete theory that will produce this type of potential.

In the second part of the paper we analyze a negatively curved field-space metric and a family of orbital-inflation potentials. These are two-field models with a light adiabatic perturbation. The effective entropic mass is large and negative on subhorizon scales and massless on superhorizon scales. As a result, the subhorizon entropic modes source an exponential growth of the adiabatic perturbation. This bounds the turning rate from above in order to keep the mass scale of inflation compatible with nucleosynthesis bounds. The entropic perturbations freeze after horizon crossing while the adiabatic perturbation grows linearly with time. Tensor modes are exponentially suppressed in this model. In addition to terminating after a sufficient number of e-folds, the trajectory's field excursion is easily made sub-Planckian. Furthermore, the potential always has either $\epsilon_V > 1$ or $\eta_V < -1$, thus globally satisfying the refined de Sitter swampland conjecture. This constitutes a previously unexamined model that satisfies the conjectures while achieving prolonged, finite, and phenomenologically viable inflation.

\section{Acknowledgements}

It is a pleasure to thank Jacques Distler for suggesting helix-like potentials as possible realizations of high-slope inflation and Irene Valenzuela for interesting discussions. We would also like to thank Gonzalo Palma for reviewing an earlier draft of this manuscript and Ana Ach\'{u}carro,  P. Christodoulidis and D. Roest for comments on the first version of the paper. This work was supported by the U.S. National Science Foundation under Grants PHY-1521186 and PHY-1620610.  This work was initiated at Aspen Center for Physics, which is supported by U.S. National Science Foundation grant PHY-1607611. 

\appendix

\section{Steady-state solution to helical potential}
\label{sec:helix_soln}

We look for a solution to the equations of motion (\ref{equ:helixeom}) with 
\begin{equation}
\begin{aligned}
z^\prime &= -\frac{1}{R}\frac{1}{1+\frac{A^2}{f^2}}\\
\theta &= z/f + c\\
\delta r &= b\, e^{z/R}
\end{aligned}
\label{eq:sseom}
\end{equation}
 where $b,c$ are constants. Near the center of the track, $b$ is small and we neglect $\mathcal{O}(b^2)$, or $\mathcal{O}(b)$ compared to constant terms in the equations of motion. In addition we neglect the small $z$-dependence in $b$ and $c$, since it is $\mathcal{O}(A^2 f^2)$ and we are interested in regime with $A$ and $f$ both small. Our solution ansatz solves the equations of motion when
\begin{align}
b &= \frac{A f \sigma^2 \csc(c)}{(A^2+f^2)R \Delta} \\
\tan{c} &= \frac{6 R^2 \left(A^2+f^2\right)-f^2}{2 f R}
\end{align}

Numerically this solution is stable in a narrow basin of attraction. Small perturbations around this solution $\delta r = b \, e^{z/R} + \delta \delta r$ are stable when the initial perturbation $\delta\delta r(t_0) \lesssim \sigma/4$. With larger perturbations, either the fields exit the track, or the metastable solution of Figures \ref{fig:metaVsSteady} and \ref{fig:helixTanhBackground} is possible. The metastable solution does eventually converge to this steady-state solution; the metastable phase in Figure \ref{fig:metaVsSteady} lasted $\sim 250$ e-folds before doing so. However in Figure \ref{fig:helixTanhBackground}, we ended inflation during the metastable phase.

\begin{figure}[h]
\centering
\includegraphics[width=0.7\textwidth]{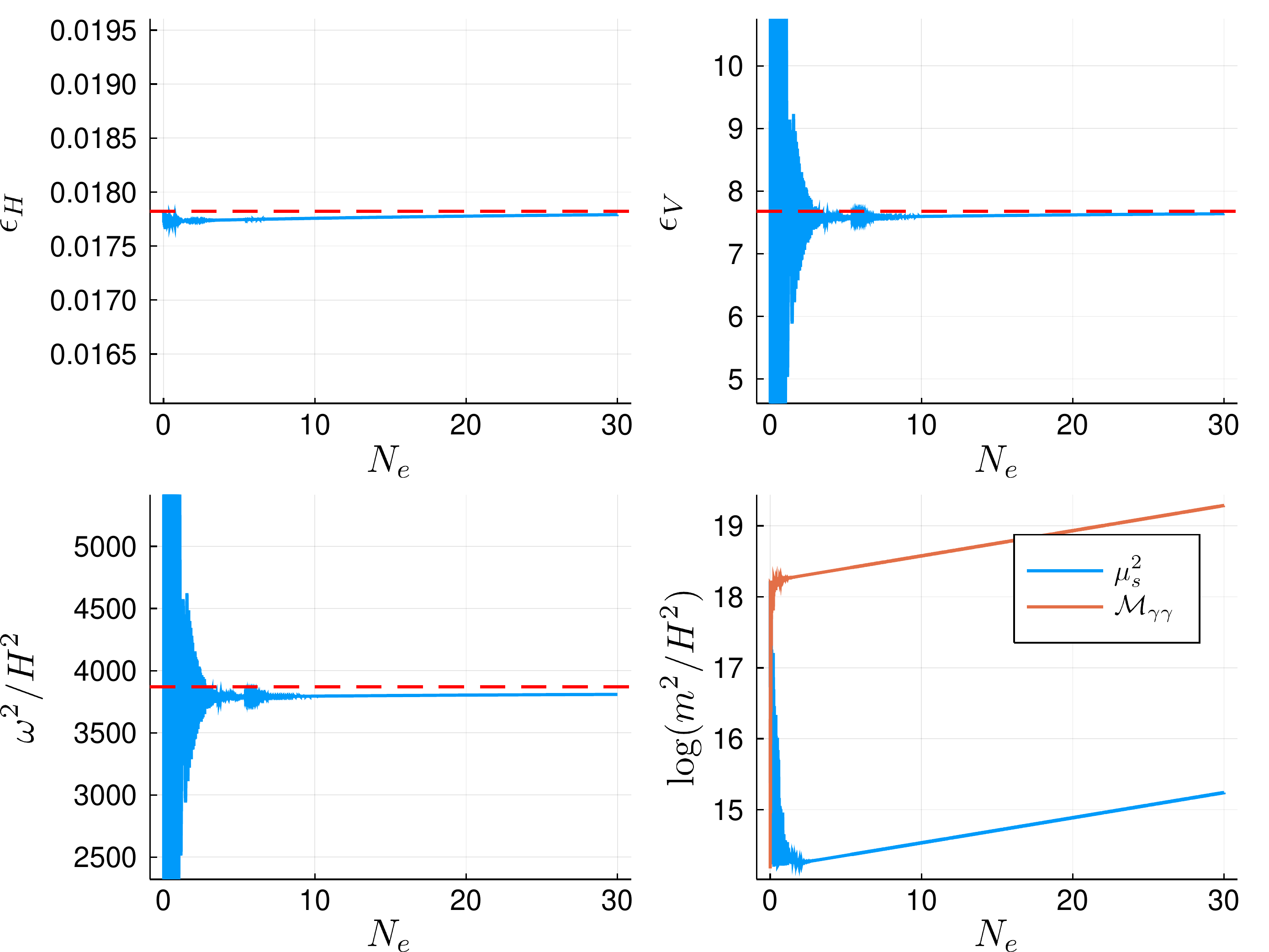}
\caption{Slow-roll parameters during the numerical evolution of our helical track potential. The numerical slow-roll parameters are the blue solid lines, and our corresponding steady-state solutions (\ref{eq:sssr}) are red dotted lines. The deviations from the steady state solution decay proportional to $e^{z/R}$. Initial conditions were chosen slightly off (\ref{eq:sseom}), which give rise to oscillations as the solution settles into the steady state within a few e-folds. The potential parameters used were $A=3\times10^{-3}\Mpl$, $f=4\times10^{-4} \Mpl$, $\Delta=2.0$, $R=0.7\Mpl$, $\sigma=10^{-3}\Mpl$. The field-space excursion in this simulation was $\sim 0.7 \Mpl$ over $30$ e-folds.}
\label{fig:helixsr}
\end{figure}

The interesting slow-roll parameters are all constants in the steady state:
\begin{equation}
\begin{aligned}
\epsilon_H &= \Mpl^2\frac{1}{2 R^2} \frac{1}{1+A^2/f^2}\\
\epsilon_V &= \Mpl^2\frac{f^2}{2(A^2+f^2)^2}\left(\frac{A^2+f^2}{R^2}+\frac{4 A^2 f^2}{(f^2 - 6(A^2+f^2)R^2)^2}\right)\\
1+\frac{\omega^2}{9 H^2} &= \epsilon_V/\epsilon_H = 1+\frac{4 A^2 f^2 R^2}{(A^2+f^2)(f^2-6(A^2+f^2)R^2)^2}.
\end{aligned}
\label{eq:sssr}
\end{equation}
This solution matches our numerics well, see figure \ref{fig:helixsr}.

This solution has high-slope inflation ($\epsilon_V \gg \epsilon_H$) in a large region of parameter space, provided we make $A$ and $f$ both small. The global properties of the potential remain identical to the discussion in section \ref{sec:helix}.
In short, the potential satisfies the refined dS conjecture locally and regions that violate the conjecture are asymptotically far, where the distance conjecture gaurantees the effective field theory is invalid anyway.

In this solution, the slow-roll parameters are constant and inflation cannot end. For the simulations in this appendix, we terminated inflation manually once it had become apparent that the perturbations had frozen on superhorizon scales, and were relatively insensitive to the end of inflation.
In our more careful analysis of section \ref{sec:helix}, our simulations were terminated by inflation ending due to a modification of the potential. 

Similarly to the metastable solution in figure \ref{fig:helixTanhBackground}, in figure \ref{fig:helixsr} the steady-state solution has both entropic masses large during the entire inflationary trajectory, so we expect an effective single-field description to be approximately valid.

In the steady state, taking the limit $b\rightarrow 0$, the speed of sound is
\begin{equation}
\begin{aligned}
c_s &= \frac{C + \left(2 A^4+2 A^2 f^2 \left(2-3 R^2\right)+3 f^4 \left(1-2 R^2\right)\right) e^{z/R}}{C + \left(2 A^4+2 A^2 f^2 \left(9 R^2+2\right)+f^4 \left(18 R^2-1\right)\right) e^{z/R}}, \\
C & \equiv \frac{2 \Delta  \Lambda ^4 R^2 \left(A^2+f^2\right)^4-2 f^4 \sigma ^2 \left(A^2+f^2\right)}{A^2 f^2 \Lambda ^4 \sigma ^2}
\end{aligned}
\end{equation}
The speed of sound can substantially differ from $1$ when $C$ is subdominant to the $e^{z/R}$ terms. In the high $z/R$ limit, $c_s$ is minimized by a small ratio of $A/f$.
This expression qualitatively agrees with our numerical calculations of (\ref{eq:cs}). For the simulation in figure \ref{fig:helixsr}, $c_s \simeq 1$, but it can be slightly lower\footnote{Parameters which provide $c_s \sim 0.8$, $n_s \sim 0.96$ are $z_0=1.0 \Mpl$,$R = 0.7 \Mpl$,$\Delta = 1.0$,$A=6\times 10^{-4}\Mpl$,$f=8 \times 10^{-5} \Mpl$,$\sigma = 1.3 \times 10^{-3} \Mpl$. In the steady state, this solution has $\omega^2/H^2 \sim 10^5, \epsilon_V \sim 190, \epsilon_H \sim 0.02$.}.

\begin{figure}[h]
\centering
\includegraphics[width=0.9\textwidth]{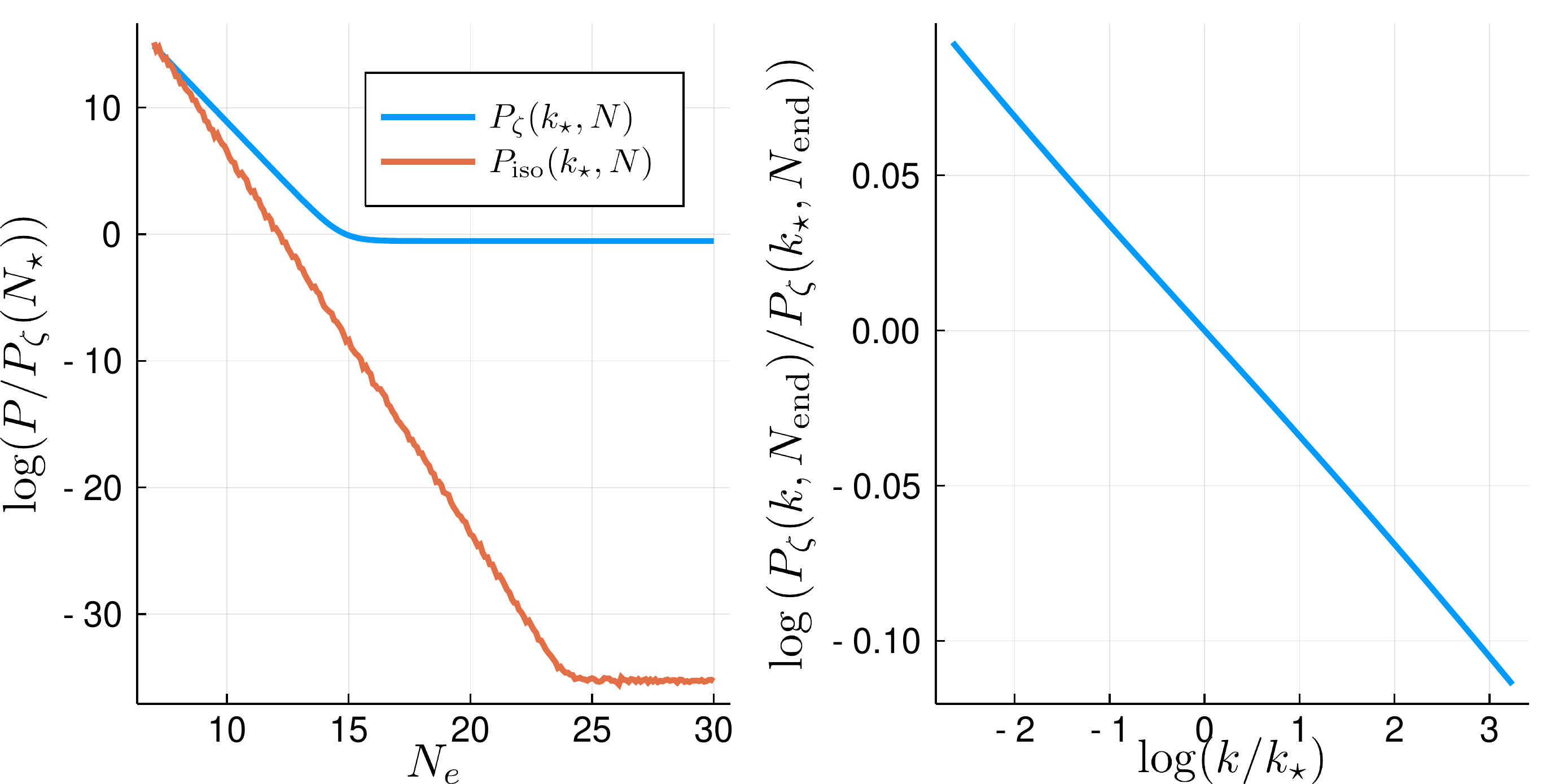}
\caption{ (Left) The powerspectra for a mode that exited the horizon $15$ e-folds after the beginning of inflation. We begin the plot when we numerically imposed Bunch-Davies initial conditions, $8$ e-folds before this mode exited the horizon. The adiabatic mode freezes on superhorizon scales, while the isocurvature powerspectra decay until they are numerically indistinguishable from zero. (Right) The adiabatic powerspectrum is smooth and featureless in $k$, with an $n_s=0.9653$. Potential parameters match those used in figure \ref{fig:helixsr}.}
\label{fig:helixpowerspectra}
\end{figure}

The spectra tilt in reduced speed-of-sound models is given in (\ref{eq:singleFieldns}), which we restate here for clarity:
\begin{align}
n_s - 1 \simeq -2 \epsilon_H - \eta_H - \kappa
\end{align}
In the steady state, $\eta_H \approx 0$ and $\kappa$ is negligibly small except for the short window of time when $C$ and the $e^{z/R}$ term are comparable in size. This region of time was avoided in our simulations. The adiabatic mode, then, influences $n_s$ only by the effects of $\epsilon_H$. Recalling our steady-state expression (\ref{eq:sssr}), we expect $\epsilon_H$ and therefore $n_s$ to be set by the ratio $A/f$. Fortunately, this is consistent with our high-slope inflation requirement, which only needs $A$ and $f$ both small. If we take $A/f \sim 7$ and $R \sim \sqrt{2}\Mpl$, then we expect $n_s\sim0.96$.
 A simulation with Planck-compatible scalar powerspectra is shown in figure \ref{fig:helixpowerspectra}.

The steady-state dynamics give low isocurvature ($r_\textrm{iso} \equiv P_\mathrm{iso}(k_\star,N_\textrm{end})/P_\zeta(k_\star,N_\textrm{end}) \sim 0$ within machine precision) and a featureless adiabatic powerspectrum. The scalar powerspectra have frozen out, so we expect these predictions to be largely independent of any late-time modification to the potential to end inflation. 

For the steady-state evolution, we did not perform any numerical analysis of the tensor perturbations or bispectrum of scalar perturbations, but applied the single-field EFT to estimate these quantities.

Recalling the single-field EFT results in equs. (\ref{eq:singleFieldns})-(\ref{eq:singleFieldfnl}), the $\epsilon_H$ in figure \ref{fig:helixsr} and a $c_s\sim 1$ gives $r\sim 0.32$ and $f_\mathrm{NL}^\mathrm{equ} \sim 0$.

Because this solution's slow-roll parameters are constants, the only way to achieve a lower $r$ is to lower $\epsilon_H$ and therefore raise $n_s$.
This solution's predictions then lie on a line in the $n_s$-$r$ plane, excluded from the Planck region at $\gtrsim 7\sigma$.

\section{Transport Method}
\label{sec:transport}

The transport method \cite{Dias:2015rca} is a robust and numerically stable technique for evolving inflationary perturbations. For convenience, we briefly summarize the method here. In this section lowercase latin indices will run from $0,\ldots,2 N_f-1$, while uppercase latin indices will be consistent with the rest of this paper and run from $1,\ldots,N_f$.

In the transport method, rather than evolving the perturbations directly, we evolve their two-point functions. For convenience we define a concatenation of the field and momenta perturbations $X^a \equiv \{Q,\delta \pi\}$, where
\begin{align}
\delta \pi^I \equiv \partial_N Q^I
\end{align}
The perturbations' equations of motion (\ref{eq:perteom}) can be written, to tree level, as
\begin{align}
\partial_N X^a = u^a_b X^b + \ldots
\end{align}
where 
\begin{align}
u^a_b \equiv \begin{pmatrix}
0 & \delta^{A}_{\bar{B}} \\
-\delta^{\bar{A}}_{B} \frac{k^2}{a^2 H^2} - \frac{\cM^{\bar{A}}_{B}}{H^2} & \delta^{\bar{A}}_{\bar{B}}(\epsilon_H-3)
\end{pmatrix}
\end{align}
We define the two-point function as
\begin{align}
\langle X^a(\vec{k}) X^b(\vec{k}^\prime) \rangle = \frac{(2\pi)^3}{k^3} \delta(\vec{k}+\vec{k}^\prime) \Sigma^{ab}.
\end{align}
We can evolve the dimensionless two-point function $\Sigma^{ab}$ in time as
\begin{align}
\Sigma^{ab}(N) &= \Gamma^a_c(N,N_0)\Gamma^b_d(N,N_0) \Sigma^{cd}(N_0)\\
D_N \Gamma^a_b &= u^a_c \Gamma^c_b
\end{align}
where $D_N$ is a covariant e-fold derivative and $\Gamma^a_b(N,N_0)$ propagates the evolution from a time with known initial conditions $N_0$, to a later time $N$.

When a mode is sufficiently subhorizon, $\Sigma^{ab}$ will be approximately the dimensionless two-point function of a Bunch-Davies state. For a mode with wavenumber $k$, this is
\begin{align}
\Sigma^{ab}\rvert_\mathrm{BD} = \begin{pmatrix}
\frac{H^2 \cG^{IJ}}{2}|k \tau|^2 &  -\frac{H^2 \cG^{\bar{I} J}}{2}|k \tau|^2 \\
-\frac{H^2 \cG^{I \bar{J}}}{2}|k \tau|^2 & \frac{H^2 \cG^{\bar{I} \bar{J}}}{2}|k \tau|^4 \\
\end{pmatrix}
\end{align}
where $k \tau = -k/(a H)$.
In our simulations, we impose these initial conditions 8 e-folds before the mode exits the horizon. Note that these initial conditions have corrections proportional to powers of $\epsilon_H$ (see around (3.9) of \cite{Dias:2015rca}). In our high-slope inflation models, $\epsilon_H$ is small at the time we impose these initial conditions.

In order to compute the physical gauge-invariant quantity $\zeta$, the adiabatic perturbation on surfaces of constant density, we need to transform out of spatially flat gauge. The relevant transformation is \cite{Dias:2015rca}
\begin{align}
N_a = \left( \frac{\pi_A}{2 \epsilon_H},0 \right)
\end{align}
where $\pi^A \equiv \partial_N \phi^A$. We can then define the $\zeta$ powerspectrum 
\begin{align}
P_\zeta(k,N) = \frac{1}{2\pi^2} N_a(N) N_b(N) \Sigma^{ab}(k,N)
\end{align}
The scalar spectral index is then
\begin{align}
n_s -1 \equiv \frac{d \log{P_\zeta} }{d \log{k}}
\end{align}
which we fit numerically, after solving for $\Sigma^{ab}$ over a range of $k$-values.
The isocurvature powerspectra are given by scalar fluctuations perpendicular to the adiabatic direction. We label a basis for these directions (i.e. the null space of $N_A$) as $v^\alpha_I$, where $\alpha$ labels the $N_f-1$ basis vectors and $I$ labels the $N_f$ vector components.
We define the isocurvature powerspectra as 
\begin{align}
P_\textrm{iso}^{\alpha \beta} &= \frac{1}{2 \pi^2} \frac{1}{2 \epsilon_H} v^\alpha_I(N) v^\beta_J(N) \Sigma^{IJ}(k,N)
\end{align}
Note that we only index the field-field quadrant of the 2-point correlation matrix for this expression. The equivalent gauge transformation here is the $2\epsilon_H$ in the denominator. In practice, we often do not care about the individual isocurvature powerspectra, but only the total amount of isocurvature. This is given by the trace $P_\mathrm{iso} \equiv \delta_{\alpha \beta} P_\mathrm{iso}^{\alpha \beta}$.

Tensor perturbations can be treated similarly to the scalar ones. For a comprehensive treatment, we again refer the reader to \cite{Dias:2015rca}.
In short, there is a two-component vector $Y^a_s = \{ \gamma_s , \pi_s \} $, where $\gamma_s$ is a scalar component of a tensor perturbation and $\pi_s \equiv \partial_N \gamma_s $ its momentum.
  The polarization is labelled by $s \in \{ +, \times \}$.
Their two-point function can be written
\begin{align}
\langle Y^a_s(\vec{k}) Y^b_{s^\prime}(\vec{k}^\prime) \rangle \equiv (2\pi)^3 \delta_{s s^\prime} \delta(\vec{k}+\vec{k}^\prime) \Upsilon^{ab}(k)
\end{align}
Similarly, we can evolve the dimensionless two-point function as
\begin{align}
\partial_N \Upsilon^{ab} = w^a_c \Upsilon^{cb} + w^b_c \Upsilon^{ac}
\end{align}
where
\begin{align}
w^a_b \equiv \begin{pmatrix}
0 & 1 \\
-k^2/(a H)^2 & \epsilon_H-3
\end{pmatrix}.
\end{align}
The corresponding initial conditions are 
\begin{align}
\Upsilon^{ab}|_\mathrm{BD} = H^2 \begin{pmatrix}
|k \tau|^2 & -|k \tau|^2  \\
-|k \tau|^2 & |k \tau|^4 
\end{pmatrix}
\end{align}
At the end of inflation, the tensor amplitude at the pivot scale is $ A_T = 4 \Upsilon^{00}(k_\star,N_\mathrm{end}) / (2\pi^2)$, and the tensor-to-scalar ratio is $r \equiv A_T / P_\zeta (k_\star,N_\mathrm{end})$.

\section{Derivation of Superpotential Model}\label{SuperDerivation}
Potentials of the form (\ref{cp_potential}) lead to the equation:

\begin{equation}
	\dot{\phi^I}=-2\cG_{IJ} \, \frac{ \partial W}{\partial \phi^J}
\end{equation}
In the two field case we considered in this paper where $W$ only depends on $X$, this implies $\dot{Y}=0$. In this Appendix we show that the converse is also true. We show that the potential (\ref{cp_potential}) may be obtained by examining $\dot{Y}=0$ solutions to (\ref{eq:multiFieldEoM}). Imposing this constraint, the equations of motion become
\begin{gather}
3H^2 = \frac{1}{2}e^{2Y/R_0}\dot{X}^2 + V \\
\ddot{X} + 3H\dot{X} + e^{-2Y/R_0}\partial_X V = 0 \\
-\frac{1}{R_0}e^{2Y/R_0}\dot{X}^2 + \partial_Y V = 0.
\end{gather}
Solving for $\dot{X}$ and taking another time derivative, we obtain
\begin{align}
\ddot{X} = \frac{R_0}{2}e^{-2Y/R_0}V_{YX}.
\end{align}
Substituting into the Friedmann equation, we find
\begin{align}
3H = \pm\sqrt{3}\sqrt{\frac{R_0}{2}V_Y + V}.
\end{align}
The $X$ equation of motion thus becomes
\begin{align}\label{eom1constY}
V_X + \frac{R_0}{2}V_{YX} = \pm\sqrt{3}e^{Y/R_0}\sqrt{V + \frac{R_0}{2}V_Y}\sqrt{R_0 V_Y},
\end{align}
or
\begin{align}\label{eom2constY}
2\partial_X\left(\sqrt{V+\frac{R_0}{2}V_Y}\right) = \pm\sqrt{3R_0}e^{Y/R_0}\sqrt{V_Y}.
\end{align}
Chen et al.'s solution arises from choosing a potential of the form $V(X,Y) = h(X) + f(X)g(Y)$. Equation (\ref{eom1constY}) then becomes
\begin{align*}
h'(X) + f'(X)g(Y) + \frac{R_0}{2}f'(X)g'(Y) = \pm\sqrt{3R_0}e^{Y/R_0}&\sqrt{h(X) + f(X)g(Y) + \frac{R_0}{2}f(X)g'(Y)} \\ \times &\sqrt{f(X)g'(Y)}.
\end{align*}
Choosing $g(Y) = -2e^{-2Y/R_0}$, this simplifies to
\begin{align}
h'(X) = \pm\sqrt{12h(X)f(X)}.
\end{align}
Defining $h(X) \equiv H^2(X)$ and $f(X) \equiv F^2(X)$, we see that
\begin{align}
[H'(X)]^2 = 3F^2(X).
\end{align}
For $W(X) \equiv \frac{H(X)}{\sqrt{3}}$, we recover the potential (\ref{cp_potential}).

A separable potential, $V(X,Y) = f(X)g(Y)$, corresponds to taking $h(X)=0$ above. Equation \eqref{eom1constY} becomes
\begin{align}
\frac{f'(X)}{\pm\sqrt{3R_0}f(X)} = \sqrt{\frac{g'(Y)e^{2Y/R_0}}{g(Y) + \frac{R_0}{2}g'(Y)}} \equiv C,
\end{align}
where $C$ is a constant. Solving for $f$ and $g$, we find
\begin{gather}
f(X) \propto e^{\pm\sqrt{3R_0}CX} \\
g(Y) \propto \exp\left[\log(2e^{2Y/R_0} - R_0 C^2) - \frac{2Y}{R_0} \right].
\end{gather}
This yields
\begin{align}
V(X,Y) = Be^{\pm\sqrt{3R_0C^2}X}\left(1 - \frac{R_0}{2}C^2 e^{-2Y/R_0} \right),
\end{align}
where $B$ is a constant.
Note that this separable potential is equivalent to the (\ref{cp_potential}) with $W(X) = Ae^{X/R'}$, $R_0 C^2 = \frac{4}{3(R')^2}$, and $B = 3A^2$.

\section{Derivation of Geodesics}\label{geoDerivation}
For the field space with metric given in (\ref{cp_metric}), we solve the geodesic equation:
\begin{align}
	(\phi'')^I + \Gamma^I_{JK}(\phi')^J(\phi')^K = 0
\end{align}
where $\lambda$ parametrizes the geodesic and primes denote derivatives with respect to $\lambda$. Using the Christoffel symbols in (\ref{cp_symbols}), we obtain geodesic equations:
\begin{align}
	X'' + \frac{2}{R_0}X'Y' = 0 \\
	Y'' - \frac{1}{R_0}e^{2Y/R_0}(X')^2 = 0.
\end{align}
The $X$ equation can be expressed as:
\begin{align}
	\partial_\lambda\left( X'e^{2Y/R_0} \right) = 0 \quad\Rightarrow\quad X'e^{2Y/R_0} = C_1
\end{align}
where $C_1$ is a constant of integration. The $Y$ equation then becomes:
\begin{align}
	Y'' - \frac{C_1^2}{R_0}e^{-2Y/R_0} = 0.
\end{align}
This admits a solution of the form:
\begin{align}
	Y(\lambda) = R_0\log\left[ \frac{C_1^2 k_1 e^{\sqrt{k_1}(k_2+\lambda)/R_0} + e^{-\sqrt{k_1}(k_2+\lambda)/R_0}}{2k_1} \right],
\end{align}
where $k_1$ and $k_2$ are constants of integration. Inserting this into the $X$ equation, we have:
\begin{align}
	X' = C_1\left[ \frac{2k_1}{C_1^2 k_1 e^{\sqrt{k_1}(k_2+\lambda)/R_0} + e^{-\sqrt{k_1}(k_2+\lambda)/R_0}} \right]^2.
\end{align}
This yields:
\begin{align}
	X(\lambda) = C - \frac{2\sqrt{k_1}R_0}{C_1}\frac{1}{C_1^2 k_1 e^{2\sqrt{k_1}(k_2+\lambda)/R_0} + 1}.
\end{align}
Inverting this and using the solution for $Y$, we obtain:
\begin{align}
	Y(X) = R_0\log\left[ \frac{R_0}{(C-X)\sqrt{\frac{2\sqrt{k_1}R_0}{C_1(C-X)} - 1}} \right].
\end{align}
Defining $K = \frac{2\sqrt{k_1}R_0}{C_1}$, this simplifies to:
\begin{align}\label{App_geodesic}
	Y(X) = R_0\log\left[ \frac{R_0}{\sqrt{C-X}\sqrt{K-C+X}} \right].
\end{align}
The parameters $C$ and $K$ may be fixed such that the geodesic passes through any two points $(X_1,Y_1)$ and $(X_2,Y_2)$ such that $X_1 \neq X_2$; doing so yields
\begin{align}
	K &= \sqrt{(X_2-X_1)^2 + 2(Q_1+Q_2)+\frac{(Q_2-Q_1)^2}{(X_2-X_1)^2}} \label{geoK}\\
	C &= \frac{1}{2}\left( X_2 + X_1 + \frac{Q_2-Q_1}{X_2-X_1} + K \right) \label{geoC}\\
	Q_1 &= R_0^2 e^{-2Y_1/R_0}, \quad Q_2 = R_0^2 e^{-2Y_2/R_0}.
\end{align}
The geodesic distance between two points $(X_i,Y_i)$ and $(X_f,Y_f)$ is given by:
\begin{align}
	S = \int ds = \int_{X_i}^{X_f} \sqrt{e^{2Y/R_0} + \left( \frac{dY}{dX} \right)^2} dX. 
\end{align}
Integrating along the path given by (\ref{App_geodesic}), we have:
\begin{align}
	S &= \frac{R_0 K}{2}\int_{X_i}^{X_f} \frac{1}{(C-X)(C-K-X)} dX \nonumber \\[5 mm]
	&= \frac{R_0}{2}\log\left[ \frac{C-X_f}{C-K-X_f}\frac{C-K-X_i}{C-X_i} \right] \label{App_geolength}.
\end{align}

\bibliography{refs}{}

\bibliographystyle{JHEP}

\end{document}